%% file: main.tex
\documentclass[a4paper,twocolumn,submitted]{quantumarticle}
\pdfoutput=1
\usepackage[utf8]{inputenc}
\usepackage[english]{babel}
\usepackage[T1]{fontenc}
\usepackage{amsmath}
\usepackage{hyperref}

\usepackage{tikz}
\usepackage{booktabs}
\usepackage{qcircuit}
\usepackage{braket}
\usepackage{csquotes}

\usepackage{biblatex}
\usepackage{listings} % for code examples
\newcommand{\openqlpc}{OpenQL$_\text{PC}$}

\lstdefinestyle{python}{
  language={Python},
  basicstyle=\small\ttfamily, 
  captionpos=b,
  extendedchars=true, 
  tabsize=2, 
  columns=fixed,
  keepspaces=true,
  showstringspaces=false,
  breaklines=true,
  frame=bt,
  numbers=left,
  numberstyle=\tiny\ttfamily,
  numbersep=4pt,
  commentstyle=\itshape\color[HTML]{28A428},
  keywordstyle=\bfseries\color[HTML]{200278}, 
  stringstyle=\slshape\color{blue} 
}

\lstdefinestyle{openql}{
    language={python},
    morekeywords={gate, Param, compile, set_value, Compiler, Program}}

\lstdefinestyle{syntaxdef}{
    basicstyle=\small\ttfamily,
    numbers=none
}

\lstdefinestyle{cQASM}{
    basicstyle=\small\ttfamily,
    commentstyle=\itshape\color[HTML]{28A428},
    keywordsprefix=\%,
    keywordstyle=\slshape\color{blue},
    keywordstyle={[2]\bfseries\color[HTML]{200278}}, 
    morekeywords={[2]rz,ry,hadamard,cnot}
}
\usepackage[capitalise]{cleveref} % for \cref
\usepackage{caption} % for subfigures
\usepackage{subcaption} % for subfigures

\begin{document}

\title{Efficient parameterised compilation for hybrid quantum programming}
\author{A.M. Krol}
\email{A.M.Krol@tudelft.nl}
\author{K. Mesman}
\author{A. Sarkar}
\author{M. M\"{o}ller}
\author{Z. Al-Ars}
\affiliation{Department of Quantum \& Computer Engineering, Delft University of Technology, The Netherlands}

\maketitle

\begin{abstract}
Near term quantum devices have the potential to outperform classical computing through the use of hybrid classical-quantum algorithms such as Variational Quantum Eigensolvers. These iterative algorithms use a classical optimiser to update a parameterised quantum circuit. Each iteration, the circuit is executed on a physical quantum processor or quantum computing simulator, and the average measurement result is passed back to the classical optimiser. When many iterations are required, the whole quantum program is also recompiled many times. We have implemented explicit parameters that prevent recompilation of the whole program in quantum programming framework OpenQL, called \openqlpc, to improve the compilation and therefore total run-time of hybrid algorithms. We compare the time required for compilation and simulation of the MAXCUT algorithm in OpenQL to the same algorithm in both PyQuil and Qiskit. With the new parameters, compilation time in OpenQL is reduced considerably for the MAXCUT benchmark. When using \openqlpc, compilation of hybrid algorithms is up to two times faster than when using PyQuil or Qiskit.
\end{abstract}

\input{01_Introduction}
\input{02_Background}
\input{03_Methods}
\input{04_Results_and_discussion}

\input{05_Conclusion}

\printbibliography

\end{document}

%% file: 01_Introduction.tex
\section{Introduction}
Even though Google announced quantum supremacy in 2019~\cite{art:googlequantumsupremacy2019}, universal, fault-tolerant quantum computers are still a thing of the future. In the meantime, Noisy Intermediate-Scale Quantum (NISQ) devices like the Google Sycamore QPU have the potential to outperform classical computers in specific cases~\cite{art:characerizingquantumsupremacyBoixo2018}. 

The NISQ era means that anybody writing quantum algorithms has to contend with a limited number of qubits and a trade-off between circuit depth and error-rates. The demonstration from Google on a 53-qubit chip had a fidelity of only 0.2\%, for example~\cite{art:googlequantumsupremacy2019}. 

% Bottlenecks
Quantum algorithm development in the NISQ era is largely done with simulations of quantum devices, which are more readily available and still faster than real quantum devices for all cases where quantum supremacy has not been achieved. And many algorithms require more (interconnected) qubits than the current state-of-the-art has to offer.
Besides, simulators offer other advantages, such as access to the full state of all qubits, error-free execution, setting of specific error rates and repeatability of "random" results~\cite{art:QX2017}. 

One such area of quantum algorithm development is hybrid quantum-classical algorithms. These are expected to be the first algorithm candidates that will result in a practical application for quantum computation~\cite{art:hybridquantumclassical2021}. Current quantum computers have too few, too error-prone qubits to be sufficient to implement purely-quantum algorithms such as Shor's factorisation algorithm~\cite{art:shors} or Grover's search algorithm~\cite{art:grovers}. But with hybrid algorithms, some of the processing is done on a classical computer, so quantum circuits with less qubits and lower depth are required for the quantum device and more fine-grained error correction can be applied~\cite{art:hybridquantumclassical2021}.

\begin{figure}[h!]
    \centering
    \includegraphics[width=\linewidth]{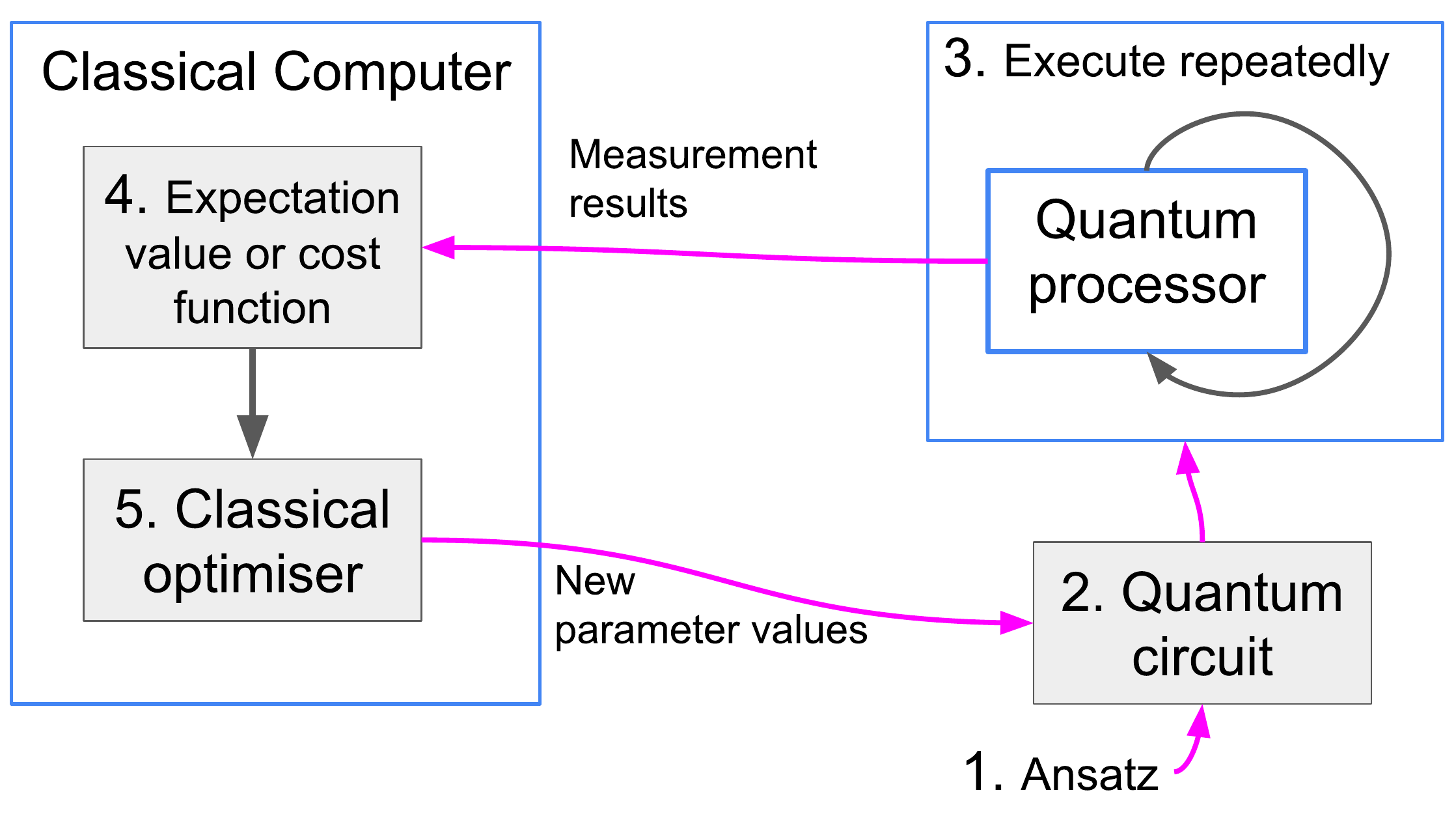}
    \caption{Programming flow for hybrid quantum algorithms like VQE}
    \label{fig:update_parameter_loop}
\end{figure}

VQE, and other variational hybrid algorithms, require many iterations of the same quantum circuit. For each iteration, a set of parameters is updated according to some classical (optimisation) algorithm~\cite{art:hybridquantumclassical2021}. An example of the program flow for such algorithms is shown in \cref{fig:update_parameter_loop}.
 
Since compilers for quantum programming languages require a lot of processing to produce an {executable} quantum circuit, doing a full compilation each iteration consumes significant amounts of time. For iterative hybrid algorithms, however, most of the circuit stays the same between iterations, which means that error-correction mechanisms, optimisations, decompositions, mapping, etc.\ are not affected. For VQE specifically, only some angles for rotation gates are changed, which means that most compilation steps are not affected for parameterised gates. 
 
With the increasing number of qubits quantum computers can support, and with increasing qubit quality, the depth and complexity of executable quantum circuits will increase. This in turn means that the amount of time spent during this recompilation step will continue to increase, making it necessary to optimise it as much as possible. %Although currently the bulk of the execution time of any hybrid algorithm is taken up by the simulator, this will not always be the case. 
At the same time, efficient classical compilation and simulation of quantum algorithms are essential tools in the NISQ era and beyond. For as long as quantum supremacy has not yet been reached, classical computers will continue to do part of the work.

%We are in the business of outperforming classical computers, and efficient compilation of quantum circuits is a part of that. 
%One aspect of that is in parameterised quantum circuits.
In this paper we introduce OpenQL Parameterized Compilation (\openqlpc) which reduces compilation time of hybrid quantum algorithms.
The contributions of this paper are as follows:
\begin{itemize}
    \item A more efficient compilation process for parameterised hybrid quantum programming
    \item Implementation of our method in the OpenQL quantum programming framework%, now called OpenQL Parameterized Compilation (\openqlpc)
    \item Implementation of the MAXCUT quantum programming benchmark in \openqlpc
\end{itemize}

This paper is structured as follows. A background on VQE and utilisation of parameters in programming languages is given in \cref{sec:background}. Then the design and implementation of parameters in OpenQL are presented in \cref{sec:designgoals,sec:syntaxandusage}, respectively. The compilation process and improvements thereof can be found in \cref{sec:compilation}. After that, the methods and results are discussed in \cref{sec:methods,sec:results}. The conclusion can be found in \cref{sec:conclusion}.

%% file: 02_Background.tex
\lstMakeShortInline[language=Python]'

\section{Background} \label{sec:background}
To demonstrate the effect that a more efficient compilation of parameters can have, the VQE algorithm will be used. We will compare our implementation, \openqlpc, against Qiskit (0.37.0)  and PyQuil (3.1.0), so some background on these will be given as well.

\subsection{Variational Quantum Eigensolvers} \label{subsec:vqe} 
VQE is a class of hybrid quantum algorithms, i.e. it uses both classical and quantum resources, to find solutions to eigenvalue and optimisation problems. With VQE, quantum devices with as few as 40-50 qubits might outperform purely classical approaches~\cite{avariationaleigenvaluesolver_Peruzzo_2014}. 
VQE can run on any gate-model quantum device, is able to leverage the strengths of a specific architecture and to variationally suppress some errors~\cite{art:theortofvariationalhybrid2016}. 
To do this, it uses a  parameterised quantum circuit, where the parameters are updated variationally according to a classical optimisation algorithm. 

% \begin{figure}[h!]
%     \centering
%     \includegraphics[width=\linewidth]{figures/param_paper_plaatjes_intro_flow.png}
%     \caption{Loop for hybrid quantum algorithms like VQE}
%     \label{fig:update_parameter_loop}
% \end{figure}

The execution flow of VQE is shown in \cref{fig:update_parameter_loop}. VQE can be used to determine the ground state and the ground state energy of a Hamiltonian $H$. A Hamiltonian is a matrix describing the total energy of a (quantum) system. VQE has the following steps~\cite{art:theortofvariationalhybrid2016}:
\begin{enumerate}
    \item Pick a parameterised quantum state that is as close as possible to the solution, in this case the ground state of the Hamiltonian. This is $\ket{\phi(\theta)}$. By adjusting the values of the parameter(s) $\theta$, the state can become (closer to) the true problem solution. This parameterised state is the ansatz.
    \item Use the ansatz and the initial parameter values to generate a parameterised executable quantum circuit that will result in the state specified by the ansatz: $\ket{0} \rightarrow \ket{\phi(\theta)}$.
    \item Repeatedly execute the circuit and measure the resulting state $\ket{\phi(\theta)}$.  This can be done by either:
    \begin{enumerate}
        \item Executing the circuit on a single quantum processor repeatedly and storing the results on a classical computer, or
        \item Executing the circuit on many quantum processors simultaneously and collecting all measurement results on a classical computer.
    \end{enumerate}
    \item Use the results from the measurements to calculate the expectation (average) value of the quantum state, or use them to calculate the energy of the state through a cost function.
    \item Use a classical optimiser to determine new parameter values for the quantum circuit that will decrease the expectation value or the cost function result.
    \item Repeat step 2-5 until the difference between the results of subsequent iterations is smaller than the desired accuracy of the result. The final parameters determine the ground state of the Hamiltonian.
\end{enumerate}

There are many options for the choice of ansatz, as well as the choice of classical optimiser. The ansatz can be tailored to many aspects of the algorithm and system used, such as the specific hardware implementation~\cite{art:theortofvariationalhybrid2016}, specific problems, accuracy or circuit depth~\cite{art:adaptivevariationalalgnature}. Different classical optimisers converge at different rates and respond differently depending on the amount of noise present in the quantum system~\cite{art:acomparisonofvariousclassical}. 

The number of iterations before algorithm convergence depends on many different details. To give an indication, the qubit efficient implementation of VQE in \cite{art:VQEwithfewerqubits} reaches the ground state in 500 iterations, with 17-qubit circuits that contain 180 to 900 variational parameters. The various VQEs from \cite{art:HWefficientVQE} are 6-qubit circuits with 30 parameters each done for 250 iterations, with $10^3$ measurements per iteration to estimate the expectation value. 

However, for the purposes of verifying and measuring \openqlpc, the specific details are not important. The choice was made to use the Nelder-Mead classical optimiser, because it is widely used and available in scipy.

\subsection{Programming of parameterised circuits} \label{subsec:bgotherlangs}
In this paper, we compare \openqlpc \ against Qiskit and PyQuil, the quantum programming languages from IBM and Rigetti, respectively. Both allow programming of parameterised circuits, are widely used and can be used as a library from a Python program. This makes them a good comparison to \openqlpc, which also has these features. 

The aim of \openqlpc \ is to be easy to use for new quantum programmers as well as for people already familiar with  these other languages. We will therefore give an overview of how parameters can be used in these other languages and aim to make our syntax similar so that adopting the new feature in OpenQL will be simple.

\subsubsection{Qiskit}
Qiskit is the quantum framework of IBM. It is suited for working with noisy qubits, which can be  simulated using their simulator, Qiskit Aer. This allows classical simulation of circuits compiled using the Qiskit compiler, Qiskit Terra. Besides simulation, circuits compiled using Qiskit Terra can also be executed on real quantum devices using IBM Q~\cite{misc:qiskitdocumentation}.

In the following code listing, we show how parameterised circuits can be used in Qiskit:
% Qiskit offers parameterised circuits like:
\begin{lstlisting}[language=Python]
qcirc = QuantumCircuit(2)
theta = Parameter("theta")
pvec = ParameterVector("pvec", 2)
qcirc.ry(theta, 0)
qcirc.crx(pvec[1], 0, 1)
qcirc.assign_parameters({pvec[1]: 1]}, inplace=True)
theta_range = np.linspace(0, 2*np.pi, 128)
circuits = [qcirc.bind_parameters({theta: theta_val}) for theta_val in theta_range]
qcirc.qasm(filename="output.qasm")
\end{lstlisting}

Defining parameters in Qiskit can be done individually, as shown above on line 2, or as a vector with a specified length, as on line 3. Both can be used as gate arguments, as on lines 4 and 5 of the example. 
Binding parameters to values can be done using the command 'assign_parameters' (line 6) to generate a single circuit, or by 'bind_parameters' to generate a list of circuits with a circuit for each value of the parameter. This is shown on line 8. This code generates a list of 128 circuits, one for each of the values of 'theta' as defined on line 7. The circuit can be output as OpenQASM to a file, as shown on line 8.

The circuits can be run on a real quantum system by compiling them individually for a specific backend. An example is shown here of how to run the circuits in the example above:
\begin{lstlisting}[language=Python]
compiled_circuit = transpile(circuits[64], simulator)
compiled_circuit = assemble(circuits, simulator)
job = simulator.run(compiled_circuit, shots=10)
counts = job.result().get_counts()
\end{lstlisting}

On line 1, one of the circuits is transpiled to be executable on the 'simulator 'backend. 
A list of circuits can also be compiled all at once, as on line 2. In both cases, the resulting 'compiled_circuit' can be executed on any supported backend with the execute command, as shown in line 3. Any measurement results can be retrieved by calling 'result().get_counts()' as on line 4. This gives the measurement results as counts of how often each possible bit combination was measured, for example: \lstinline|{'0': 4, '1': 6}|.

Compilation, execution and binding of parameter values can also be combined into a single 'execute' command, as shown below:
\begin{lstlisting}
job = execute(compiled_circuit,
              backend=QasmSimulator(),
              parameter_binds=[{theta: theta_val} for theta_val in theta_range])
\end{lstlisting}

Using a single command to bind parameters and simulate the circuit makes it impossible to determine the individual execution time of the components from the host Python program. So to get those results the separate commands are used. The compilation is considered complete after the 'assemble' command.

\subsubsection{PyQuil} 
PyQuil is the quantum programming language from Rigetti. It is compiled using the Quil compiler into Quil, also the name of their Quantum Instruction Language~\cite{misc:quilspec}.
PyQuil can be used to write hybrid algorithms with parameters~\cite{art:qcspaperrigetti}, which makes it a suitable comparison to \openqlpc. 

A PyQuil program with the same overall functionality as the one in Qiskit is shown below. The example program will be used to explain how parameters can be defined and used in PyQuil~\cite{misc:pyquilgithub}:
\begin{lstlisting}
program = Program()
ro = program.declare('ro', memory_type='BIT', memory_size=1)
theta = program.declare('theta', memory_type='REAL')
program += RX(theta, 0)
program += MEASURE(0, ro[0])
\end{lstlisting}
A quantum program is defined on line 1. Two types of parameter are declared on lines 2 and 3; 'ro' will be used to store the measurement results, and the parameter 'theta' is used as argument for the 'RY' gate on line 4. 

The parameterised quantum program 'program' from this example can be run for different values of 'theta', as shown in the following code listing~\cite{misc:pyquildoc}: 
% And executing this program for different values of theta is done like so~\cite{misc:pyquildoc}:
\begin{lstlisting}
parametric_measurements = []
executable = simulator.compile(program)
for theta_val in np.linspace(0, 2 * np.pi, 128):
    executable.write_memory(region_name='theta', value=theta_val) 
    result = simulator.run(executable)
    parametric_measurements.append(result) 
\end{lstlisting}
In line 1, the array 'parametric_measurements' is defined for storing the measurement results. The 'program' from the previous example is compiled for a specific execution platform, 'simulator 'in this case. A for-loop is used to iterate over the values of 'theta_val' (lines 3-6). On line 4, the value of 'theta_val' is written to memory region 'theta'. The program is then run on a simulator and the measurement result is appended to the list 'parametric_measurements' on lines 5 and 6.

%% file: 03_Methods.tex
\section{Design goals} \label{sec:designgoals}
Implementation of parameters in \openqlpc was done with the following ideas in mind: modularity, scalability, user-friendliness (usability), future-proof and speed.

\textbf{Modularity:} To make \openqlpc \ robust against future changes to the host programming language, the parameters will be implemented as a separate entity.

\textbf{Scalability:} This has two parts, the first is that \openqlpc \ will make it easy for the programmer to define many parameters, and to allow putting values to them in bulk as well. The second part is that the overall compilation time should not be affected (too much) by the number of parameters.

\textbf{Usability:} The syntax and use of the parameters will be made similar to other quantum programming languages (Qiskit and PyQuil), for users already familiar with those and to make manual rewriting of a program from one language to the other easier. Clear errors should be provided, and parameter types should be explicit. Finally, the programmer should not be required to manually provide a string for each parameter, which can become cumbersome for bigger programs. But manual naming should be supported, in case human-readable cQASM is desired.

\textbf{Future-proof:} Parameterisation also means a more clearly defined boundary between the static and dynamic parts of a quantum circuit, which can be used in the future for a full split between those two compiler functionalities, or for detection of parallelism. In addition, the compile speed, number of supported parameters etc.\ should allow this feature to be used (and useful) into a future of quantum programming languages, where extensive knowledge of quantum is no longer required to write quantum programs. 

\textbf{Compilation speed:} OpenQL is a high-level quantum programming language with an extensive compilation toolchain~\cite{art:khammassi2020openql}. It supports gate decomposition, circuit optimisation, scheduling and mapping, all of which are necessary to be able to execute the generated common Quantum Assembly Language (cQASM) on NISQ devices~\cite{art:proglangandcompilerdesign}. This also makes the compilation in OpenQL and all other such compilers computationally expensive, in terms of classical resources. For algorithms that require only a single compilation pass, the classical compilation time is negligible compared to execution of a quantum program on a real quantum device, or classical simulation of the circuit. However for iterative hybrid algorithms, such as VQE, repeated compilation of essentially the same quantum circuit becomes a much bigger drain on resources. With explicit definition of the dynamic parts of the circuit through the use of parameters, the bulk of the compiler operations does not need to be repeated for each iteration of the hybrid algorithm. 

\section{Syntax and usage} \label{sec:syntaxandusage}
To use a parameter, it first needs to be created, and at some point, it needs a (numerical) value. Assigning a numerical value to the parameter can be done at construction, individually at any point in the program or at compile time. All this is explained in more detail below.

\begin{table}
\centering
\caption{Parameter syntax}
\label{tab:parameter-syntax}
\begin{tabular}{@{}p{\linewidth}@{}}
\toprule \addlinespace[1ex]
$\bigl[$openql.openql.$\bigr]$ Param( \textit{Type} $\bigl[$, \textit{Name} $\bigr]$ $\bigl[$, \textit{Value} $\bigr]$ ) \\ \addlinespace[0.5ex]  \midrule \addlinespace[1ex]
~~~\textit{Type} : "INT" $|$ "REAL" $|$ "ANGLE" \\ \addlinespace[1ex]
~~~\textit{Name} : Symbolic name of the parameter, will appear unmodified in resulting cQASM (string) \\\addlinespace[1ex]
~~~\textit{Value} : Numerical value, must match the type as specified by \textit{Type} \\ \bottomrule
\end{tabular}
\end{table}

% \subsubsection*{At construction} 
\subsection{Parameter construction} \label{sec:syntaxatconstruction}
Parameter construction requires the specification of a type, which can be one of "INT", "REAL" or "ANGLE". "REAL" and "ANGLE" are essentially the same, both map to an underlying "double" type. Depending on the type, a parameter can substitute hard coded qubit numbers or gate angles. Optionally, the user can specify a name for the parameter at construction or directly assign a value to it. The syntax for parameter construction can be found in \cref{tab:parameter-syntax}. In the code listing below, example code is shown for parameter construction with specification of 1. parameter type only, 2. type and parameter name, 3. type and numerical value, and 4. type, parameter name and numerical value. 
\begin{lstlisting}[style=openql]
p_int = ql.Param("INT")
p_real = ql.Param("REAL", "pname")
p_angle = ql.Param("ANGLE", 1.724)
p_int2 = ql.Param("INT", "pname2", 4)
\end{lstlisting}
%If the parameter name is not specified, it is set to a random string of eight alphanumerical characters. There is no check whether there are duplicate parameter names, so this length was chosen to  make it very unlikely that two different randomly generated names are the same ($p=4.58^{-15}$). 
%A type error is thrown if the type of the numerical value does not match the type of the parameter, to prevent mixing of parameter types by the user.

\subsection{Using parameters} \label{sec:usingparams}
Parameters can be used in quantum circuits, in place of hard coded qubit numbers or gate angles. Some examples of using parameters are shown below, where the parameters are as defined in \cref{sec:syntaxatconstruction}:
\begin{lstlisting}[style=openql]
kernel.gate("hadamard", p_int)
kernel.gate("rz", [0], 0, p_real)
kernel.gate("ry", p_int2, p_angle)
kernel.gate("cnot", p_int, p_int2)
\end{lstlisting}
On line 1 a "hadamard" gate is added to the 'kernel', where the parameter 'p_int' is used in place of the qubit number. On line 2, parameter 'p_real' is used as the rotation angle of the "rz" gate. Both of these can be combined, as on line 3, where the "ry" gate is applied to the qubit number stored in parameter 'p_int2' and with the angle stored in 'p_angle'. Parameters can also be used for 2-qubit gates, as shown on line 4, where a 'cnot' is applied to qubits 'p_int' and 'p_int2'.

\subsection{Compiling parameters} \label{sec:compiling parameters}
Quantum programs with parameters can be compiled in the same way as circuits without parameters in OpenQL, as shown in the code listing below:
\begin{lstlisting}
compiler.compile(Program)
\end{lstlisting}
When the 'kernel' from \cref{sec:usingparams} is added to a program and compiled in this manner, the resulting cQASM output is shown below:
%Compiling circuits with parameters is the same as a circuit without parameters in OpenQL: \lstinline{c.compile(program)}. 
%This results in the following cQASM output:
\begin{lstlisting}[style=cQASM]
    hadamard %w5Bq1DRO
    rz q[0] %pname
    ry q[4], 1.724
    cnot %w5Bq1DRO, q[4]
\end{lstlisting}
Line 1 shows the 'hadamard' gate from \cref{sec:usingparams} with parameter 'p_int'. Symbolic names in cQASM are preceded by the \% symbol, and the randomly generated string "\lstinline[style=cQASM,morekeywords={w5Bq1DRO}]|w5Bq1DRO|" is the "name" of this parameter. 
Line 2 shows the 'rz' gate, applied to qubit 0. The angle is the parameter 'p_real' from \cref{sec:syntaxatconstruction}. The name of this parameter was set as 'pname' at construction, and this is reflected in the cQASM output. 
On line 3, the 'ry' gate had 'p_int2' in place of a qubit number and 'p_angle' in place of a rotation angle. Both were constructed with a numerical value already set, which is reflected in the cQASM output above, where instead of symbolic names the numerical values are used; qubit number 4 ('q[4]') for 'p_int2' and 1.724 for 'p_angle'

\subsection{Setting parameter values} \label{sec:setparamvalues}
There are three ways to set the value of a parameter in \openqlpc. These are:
\begin{itemize}
 \item at construction: \lstinline{p1 = Param(string, value)},  
 \item at any point individually:\\ \lstinline{Param.set_value(value)}, and 
 \item at compile time: \lstinline{Compiler.compile(program, [p1], [value])}
\end{itemize}

Setting a value at construction is outlined in \cref{sec:syntaxatconstruction}, examples for the other two ways are given here.
% Below, we give examples to how these three methods are used.

Assigning a numerical value to a single parameter can be done at any point in the code by calling the 'set_value' method. This is shown in the code below, where the 'p_int2' is defined as in \cref{sec:syntaxatconstruction}.
\begin{lstlisting}[style=openql]
p_int2.set_value(2)
\end{lstlisting}
This assigns the value of 2 to 'p_int'. 
It is also possible to modify the numerical value of a parameter in this way. 

When compiling and generating cQASM, the final value of a parameter will be used for all instances of a parameter, so it is not possible to assign different numerical values to a single parameter partway through a quantum circuit. The intended use is to modify parameter values between different iterations of a circuit, where the whole circuit is compiled for each iteration.

It is also possible to set values at compile time. This can be done for all parameters at once, or for a subset of the parameters. An example using the program from \cref{sec:usingparams} is shown below:
\begin{lstlisting}[style=openql]
compiler.compile(program, [p_int, p_real, p_angle], [1, 2.1, -1.7])
\end{lstlisting}
With this line of code, the values of 1, 2.1 and -1.7 are assigned to parameters 'p_int', 'p_real' and 'p_angle', respectively, and the whole program is compiled. 

This results in the following cQASM output:
\begin{lstlisting}[style=cQASM]
    hadamard q[1]
    rz q[0], 2.1
    ry q[4], -1.7
    cnot q[1], q[4]
\end{lstlisting}
On line 1, the 'hadamard' gate from before, with parameter 'p_int', is applied to qubit 1 (q[1]), the value stored in 'p_int'. This overwrites the value set in \cref{sec:setparamvalues}. The 'rz' gate on line 2 now uses an angle of 2.1, the value from 'p_real'. The 'ry' gate on line 3 is applied to qubit 4 as in \cref{sec:setparamvalues}, since the value of 'p_int2' was not modified. The angle is now '-1.7', the value assigned to 'p_angle' at compilation. On line 4, the 'cnot' gate is applied from qubit 1, as stored in 'p_int' and to qubit 4, as stored in 'p_int2'.

The resulting cQASM no longer contains symbolic names, and can now be executed on a simulator or a real quantum device.

\section{Compilation} \label{sec:compilation}
\begin{figure}[ht]
\centering
    \includegraphics[width=\linewidth]{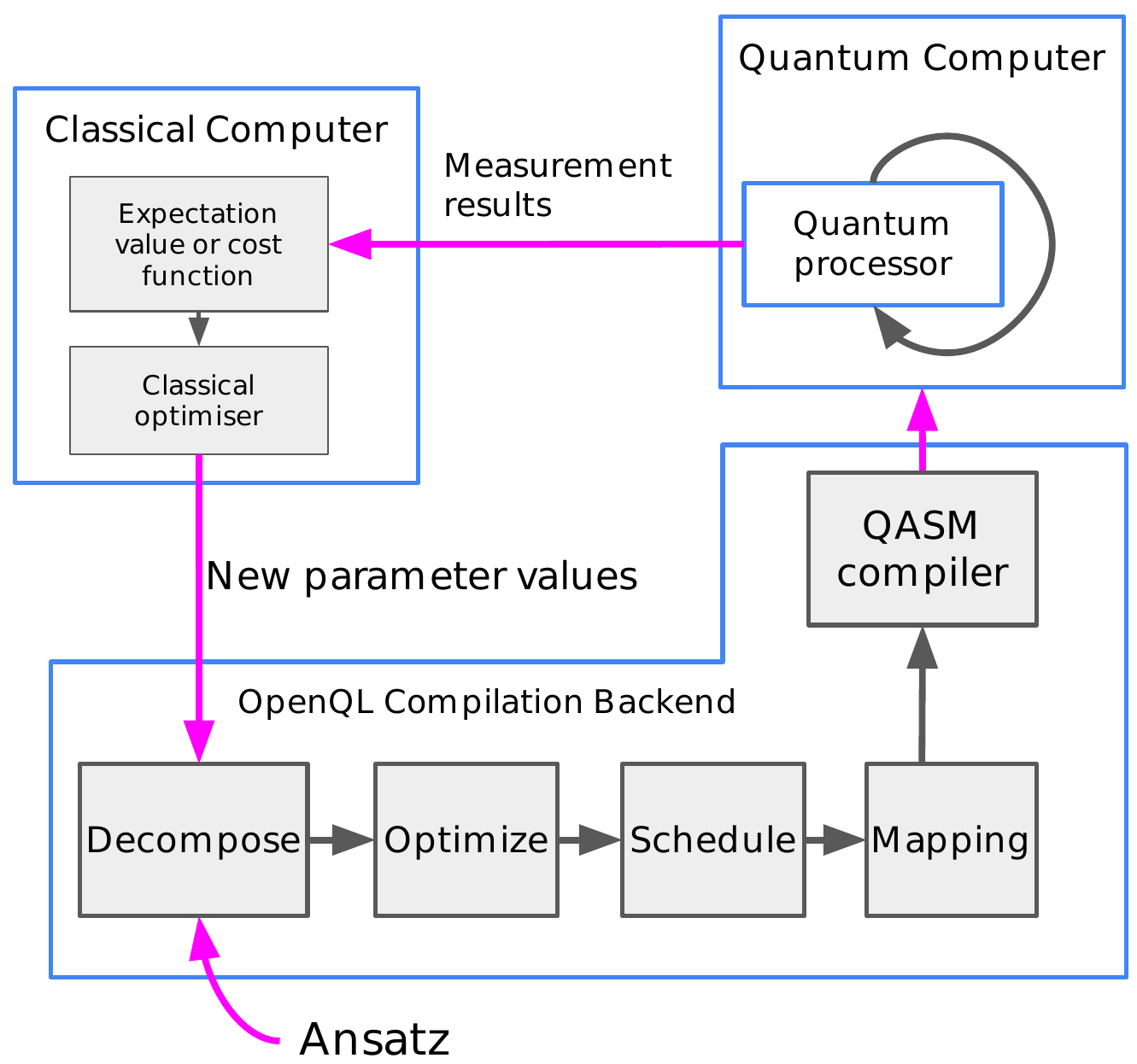}
    \caption{Compilation flow of OpenQL without using the new parameter implementation~\cite{art:khammassi2020openql,art:quaser}}
    \label{fig:parameterscompileflow_old_flow}
\end{figure}

The first time a quantum program is compiled (by calling compiler.compile(..)), the full stack is executed, including gate decompositions, mapping, optimisations, etc.~\cite{art:khammassi2020openql}. Without using our \openqlpc \ approach, updating of any (parameter) values requires repeated execution of this whole stack, as shown in \cref{fig:parameterscompileflow_old_flow}.

\subsection{Compile flow}
\begin{figure}[ht]
\centering
    \includegraphics[width=\linewidth]{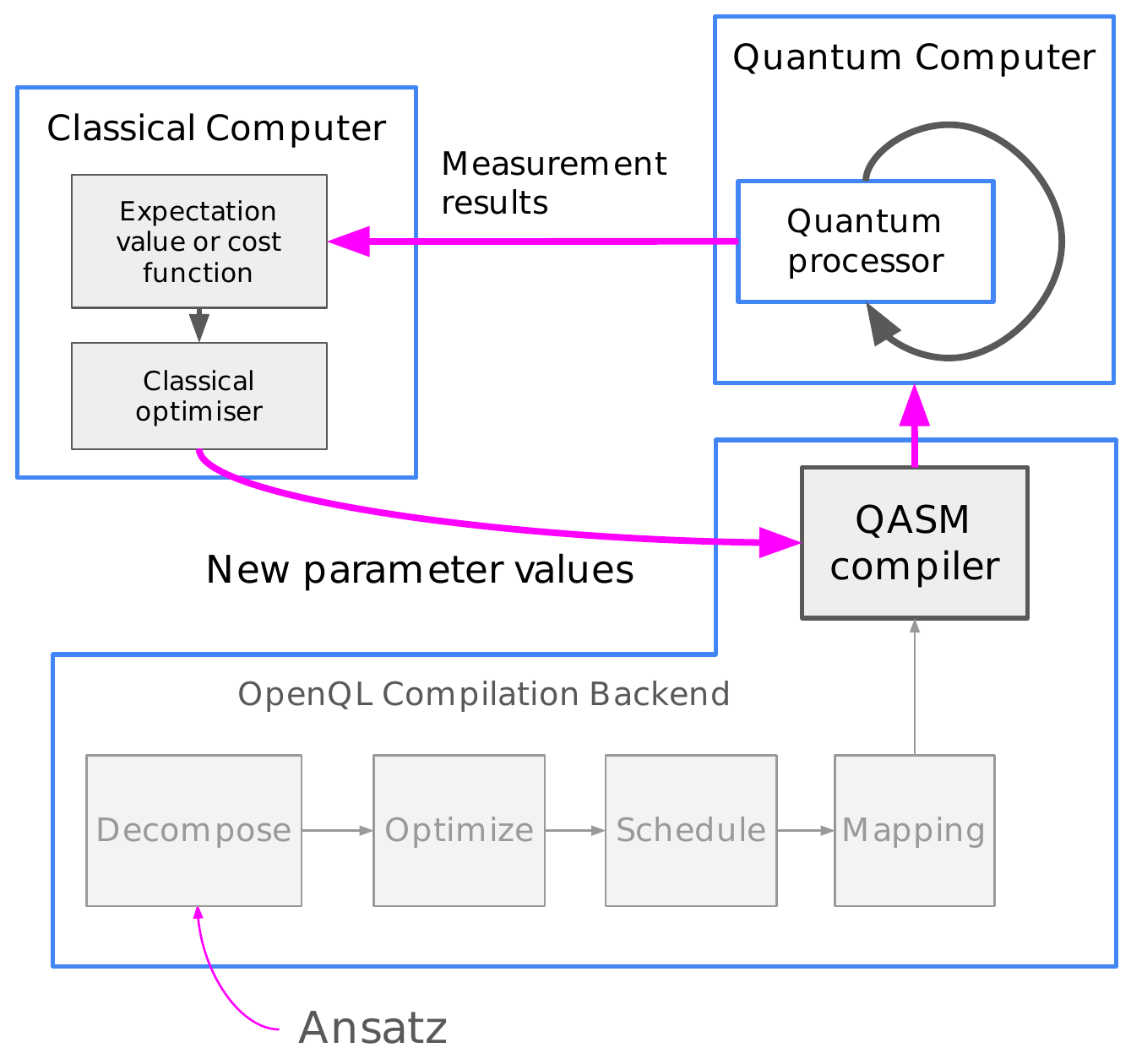}
    \caption{Compilation flow of \openqlpc \ when using the new parameter implementation}
    \label{fig:parameterscompileflow_new_flow}
\end{figure}

With \openqlpc, only the affected gates are updated with the numerical values of the parameters as shown in \cref{fig:parameterscompileflow_new_flow}. This means that there is no unnecessary repetition of the whole (extensive) compiler stack for every iteration of an iterative quantum algorithm. 


In the first compile run, any parameterised quantum instructions without a numerical value are skipped by any compiler passes that require that specific value. Mapping and scheduling one (or a set of) instruction cannot be done when the qubit is not specified, but do not require knowledge of the angle of a rotation gate.
When recompiling, only parameterised quantum instructions and affected instructions are considered, while the rest of the code is left unaffected. This might mean some optimisations that affect large regions of the code will not be done, but this effect is expected to be small compared to the resources saved  by not checking all of the code after each compilation.

\subsection{Implementation example}
An example for how parameters can be used is shown below:
\begin{lstlisting}[style=openql]
from openql import openql as ql
kernel = ql.Kernel(..)
program = ql.Program(..)
compiler = ql.Compiler(..)
kernel.gate("rz", [0], 0, theta)
program.add_kernel(kernel)
compiler.compile(program)
theta = ql.Param("ANGLE")
theta_range = np.linspace(-2*np.pi, 2*np.pi, 128)
for theta_val in theta_range:
    compiler.compile(program, [theta], [theta_val])
\end{lstlisting}
 The quantum program in this example is just a single 'rz' gate with parameterised gate angle 'theta', which is applied to qubit 0. The program is compiled in a for-loop, for the range of angles specified in 'theta_range' on line 9.
 
\begin{figure}[ht]
    \centering
    \includegraphics[width=\linewidth]{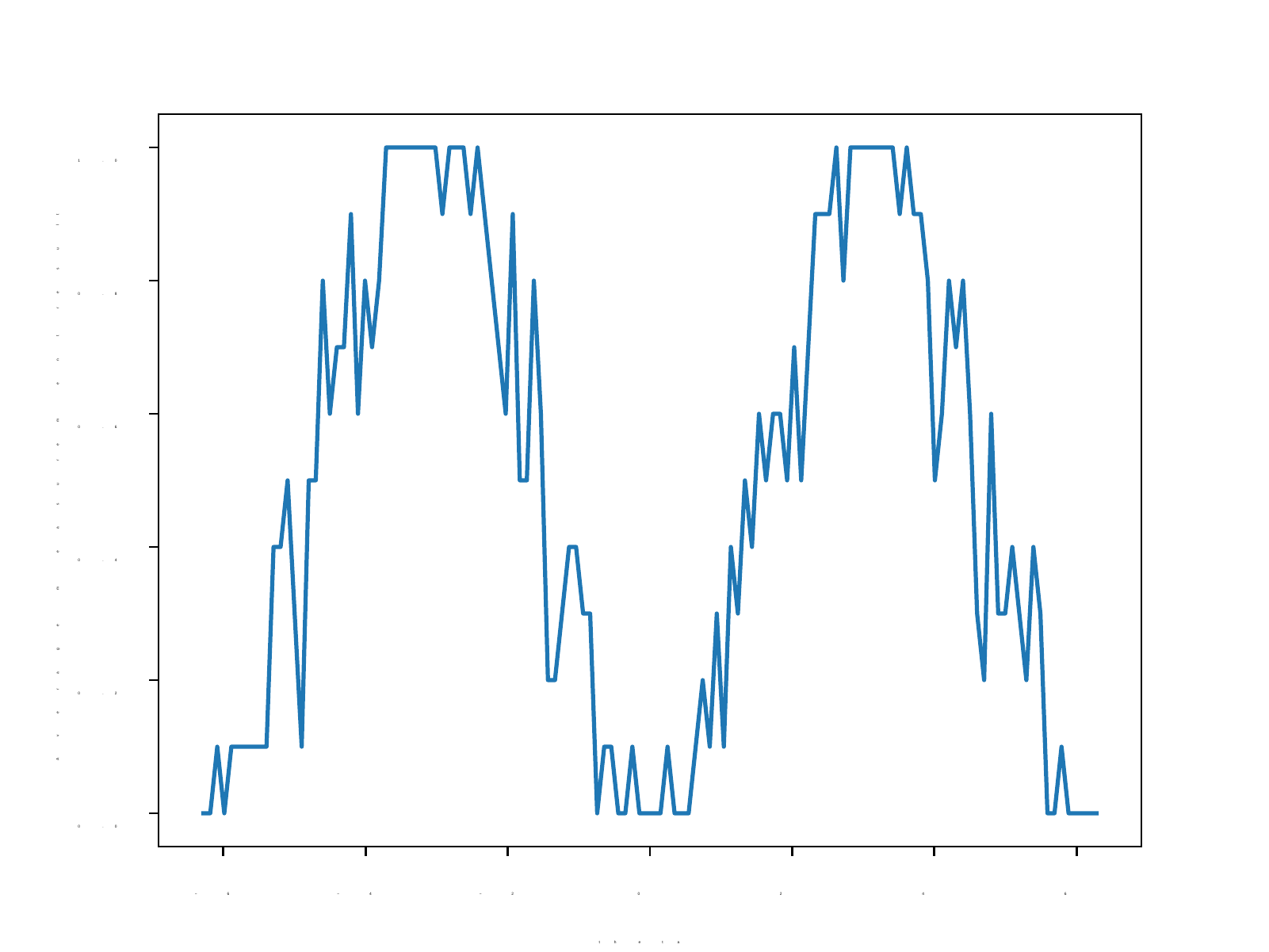}
    \caption{Average measurements results of the example \openqlpc{} program for variable angles theta}
    \label{fig:pyplot_output}
\end{figure}

When this code is run multiple times, the average measurement result for each angle theta is shown in \cref{fig:pyplot_output}.

%% file: 04_Results_and_discussion.tex
\section{Methods} \label{sec:methods}
To determine the influence of parameters, execution time tests were performed. These were done with the MAXCUT benchmark~\cite{art:qpack}. 

\subsection{The MAXCUT benchmark} \label{subsec:maxcut}
MAXCUT is a hybrid algorithm which can be used for circuit layout design~\cite{art:Barahona1988AnAO}, statistical physics and more~\cite{art:WANG2010240}. It aims to find the division of a graph into two parts, where the total weight of all edges between the parts is maximised~\cite{art:WANG2010240}. The general case is an NP-hard problem. 

\begin{figure}[ht]
    \centering
    \begin{subfigure}[c]{0.45\linewidth}
    \includegraphics[width=0.95\linewidth]{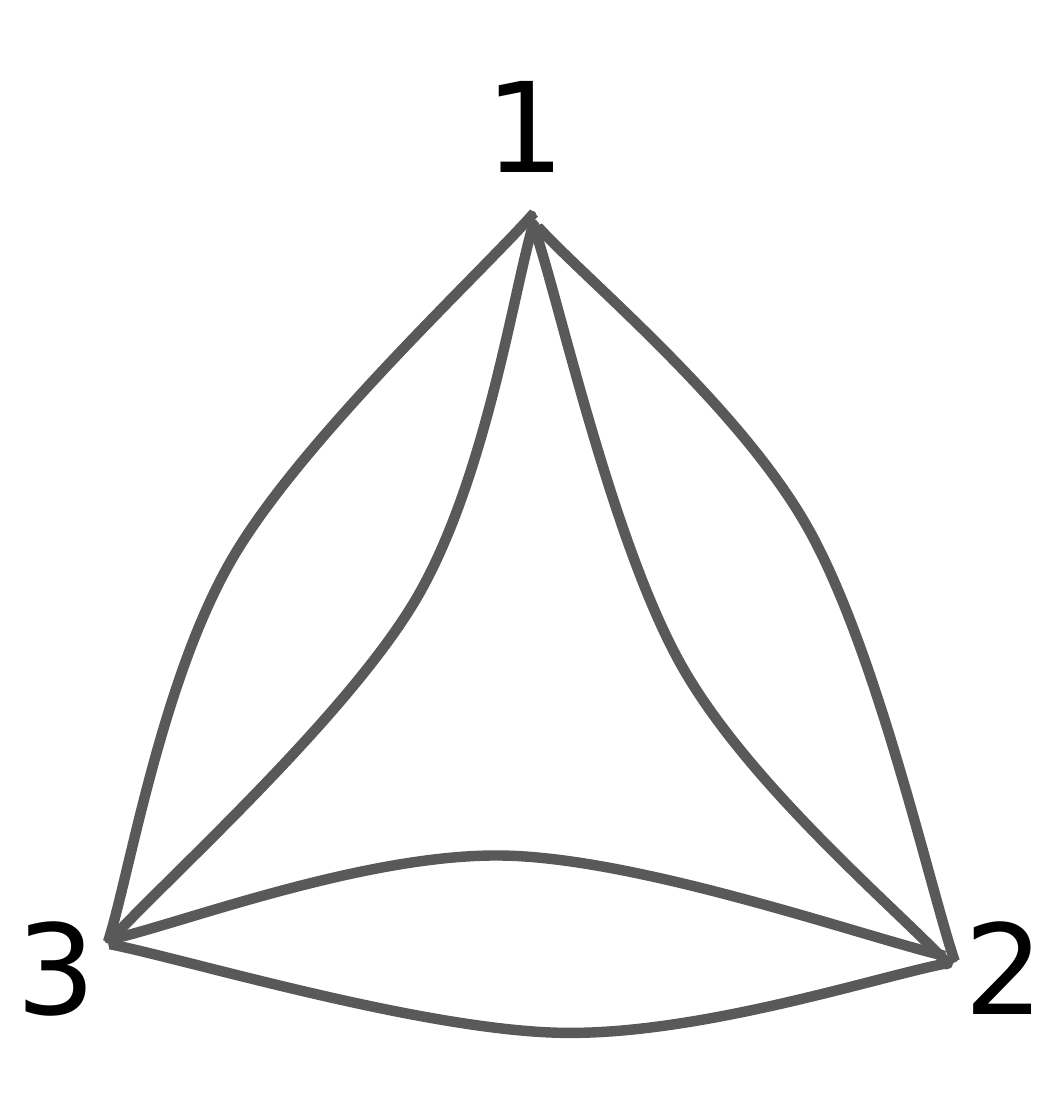}
    \end{subfigure}
    \begin{subfigure}[c]{0.45\linewidth}
    \includegraphics[width=\linewidth]{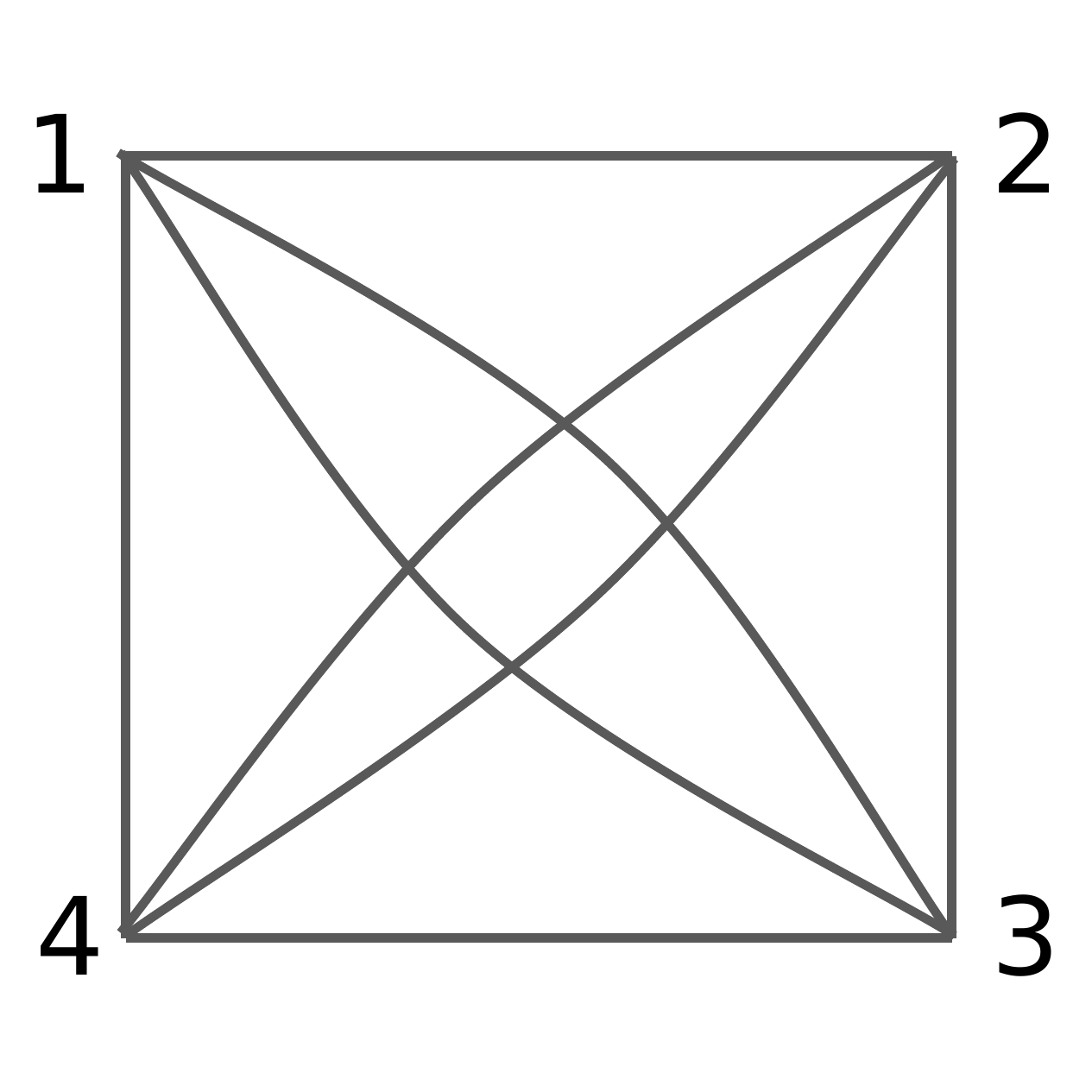}
    \end{subfigure}
    \begin{subfigure}[c]{0.49\linewidth}
    \includegraphics[width=\linewidth]{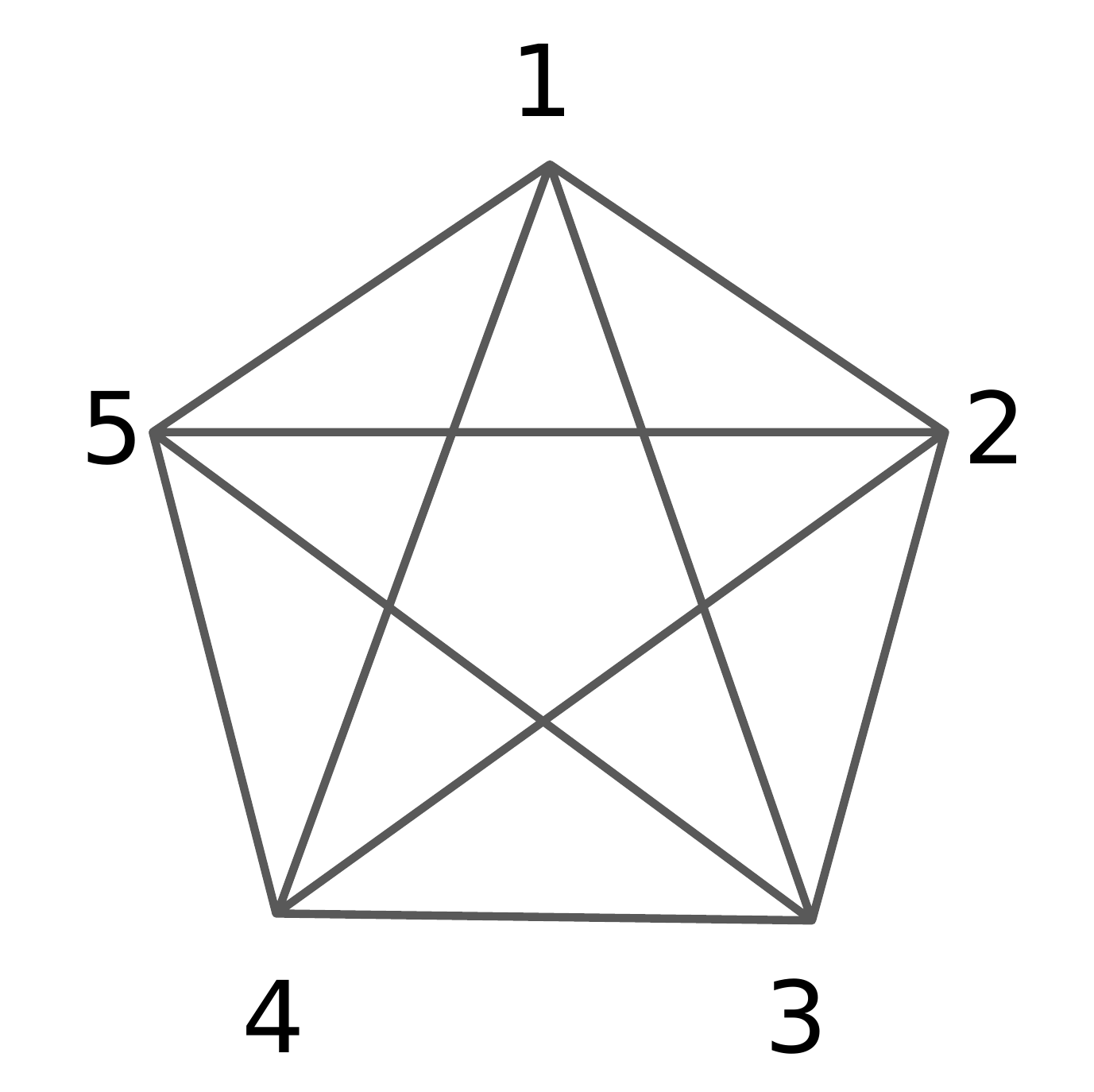}
    \end{subfigure}
    \begin{subfigure}[c]{0.49\linewidth}
    \includegraphics[width=\linewidth]{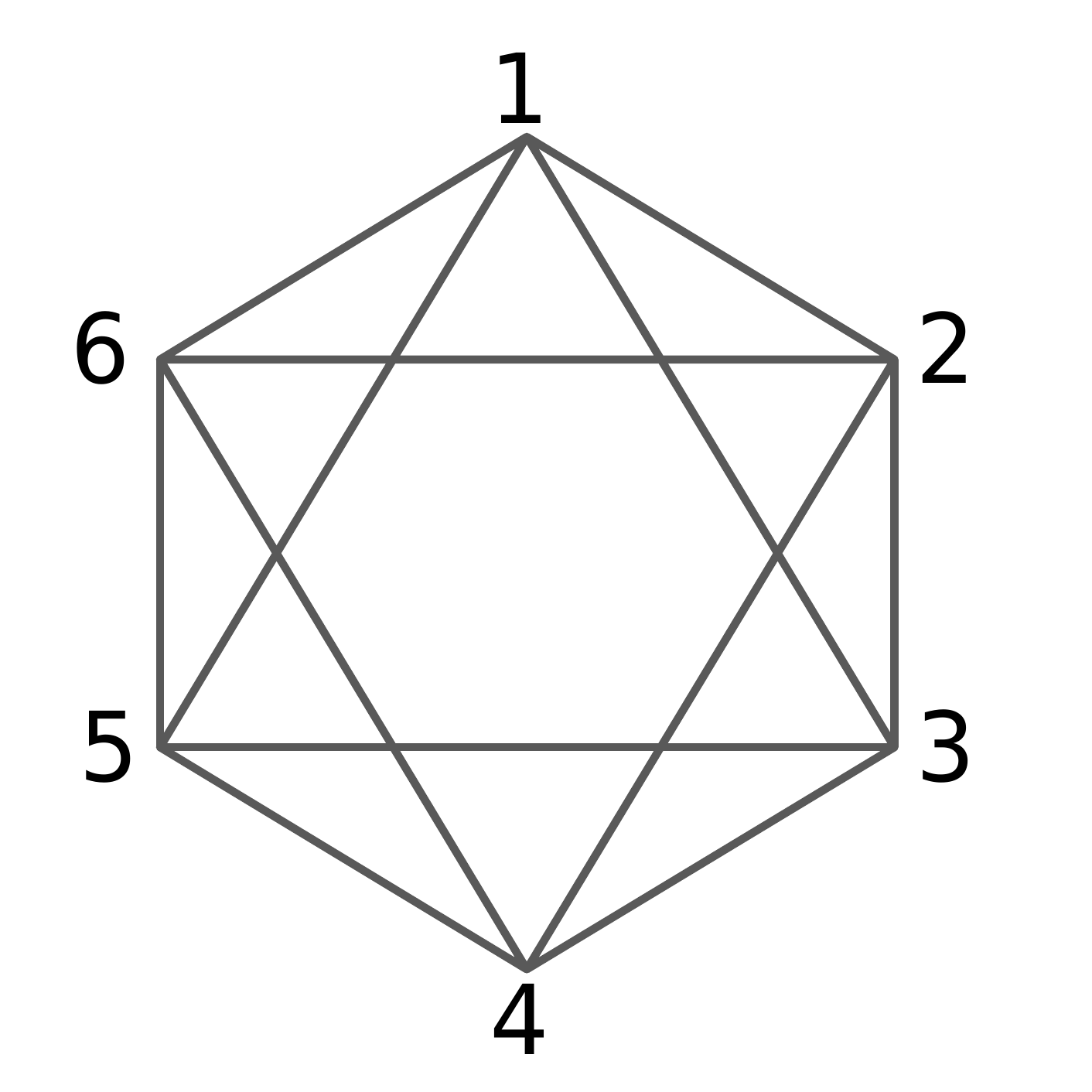}
    \end{subfigure}
    \caption{4-regular graphs with 3, 4, 5 and 6 nodes, generated as input for the MAXCUT algorithm}
    \label{fig:maxcutgraphs}
\end{figure}

The MAXCUT benchmark from~\cite{art:qpack} was implemented in OpenQL, PyQuil and Qiskit on 4-regular graphs of varying number of nodes, which are shown in \cref{fig:maxcutgraphs}. These graphs and this benchmark were chosen because a reference implementation was already available, it can be easily scaled up to any number of nodes and the number of qubits and the circuit depth both increase linearly with an increasing number of nodes. 
% These graphs were generated using Python, with the program shown below: 

% \begin{lstlisting}[language=Python]
% def regular_graph(n):
%     edges = []
%     for i in range(n-1):
%         edges.append([i, i+1])
%     edges.append([0, n-1])
%     for i in range(n-2):
%         edges.append([i, i+2])
%     edges.append([0, n-2])
%     edges.append([1, n-1])
%     return [n, edges]
% \end{lstlisting}

\lstDeleteShortInline'
% The graphs generated by this code look like the ones in \cref{fig:maxcutgraphs}, where graphs with 3, 4, 5 and 6 nodes are shown. Each node has four edges, one with each neighbour and one with its neighbour's neighbour. 

The MAXCUT benchmark is a variational quantum algorithm, with an execution flow as in \cref{subsec:vqe}. The quantum circuit for the benchmark consists of the following:
\begin{itemize}
    \item For a graph with $n$ nodes, $n$ qubits
    \item A hadamard gate on each qubit 
    \item Unitary operator $U(\gamma)$ consisting of:
    \begin{itemize}
        \item For every edge between between nodes $a$ and $b$, a CNOT gate from qubit $a$ to $b$
        \item A $R_z(\gamma)$ on qubit $b$
        \item Another cnot gate from qubit $a$ to $b$
    \end{itemize}
    \item Unitary operator $U(\beta)$ consisting of:
    \begin{itemize}
        \item A $R_x(\beta)$ gate on every qubit
    \end{itemize}
    \item Measurement of every qubit
\end{itemize}

The example below shows the circuit for the MAXCUT benchmark on a graph of two nodes, $a$ and $b$, with a single line between them. 

\begin{tabular}{c}
\Qcircuit @C=1em @R=.7em {
\push{\rule{0em}{1em}} & & & & & & & & \\
\push{\rule{0em}{1em}} & & & & & \ustick{p} & & & \\
\push{\rule{0em}{1em}} & \lstick{\ket{a}} & \gate{H} & \ctrl{1} & \qw & \ctrl{1} & \qw & \gate{R_x(\beta)} & \meter \\
\push{\rule{0em}{1em}} & \lstick{\ket{b}} & \gate{H} & \targ & \gate{R_z(\gamma)} &  \targ  & \qw & \gate{R_x(\beta)}  & \meter  \gategroup{3}{4}{4}{6}{1.2em}{--} \gategroup{3}{8}{4}{8}{1em}{--} \gategroup{3}{4}{4}{8}{2em}{^\}}\\
& & & &\dstick{U(\gamma)} & & &\dstick{U(\beta)} &  \\
\push{\rule{0em}{1em}} & & & & & & & & 
}
\end{tabular}

The unitary operators can be repeated for $p$ steps as $U(\gamma_1)U(\beta_1)U(\gamma_2)U(\beta_2) \ldots U(\gamma_p)U(\beta_p)$. Increasing the number of these steps improves the quality of the approximation \cite{art:qaoafarhi}. The total number of gates (excluding measurement) is $n\cdot(1+7\cdot p)$, for $n$ qubits and $p$ steps. For our tests, we have chosen to go with $p=3$ steps, which matches the number of steps in the benchmark. 

% Increasing the number of steps increases the number of parameters and the circuit depth linearly. An increasing number of nodes increases linearly the number of qubits and the total circuit depth, but has no influence on the number of parameters. 

\subsection{Experimental setup} \label{subsec:experimentalsetup}

% \begin{figure*}
%     \centering
%     \includegraphics[width=0.9\linewidth]{figures/openql_pc_qiskit_acc_comp_time_steps.pdf}
%     \caption{Accumulated compilation time of the MAXCUT benchmark for Qiskit, OpenQL and \openqlpc{}}
%     \label{fig:maxcut_comp_time_openql_w_wo_qiskit}
% \end{figure*}

In order to compare the performance of \openqlpc{}  with other programming languages, the MAXCUT algorithm was implemented in OpenQL and also in Qiskit (0.37.0) and PyQuil (3.1.0). 
Timing was done from the host Python program using the "time" package, on a Dell Latitude 7400 with an 8th Generation Intel Core™ i7-8665U Processor and 2x 4GiB DDR4 RAM. %For OpenQL and Qiskit, it was possible to measure the compilation time of the circuit and the simulation separately, unfortunately this was not the case for PyQuil since assigning parameter values could not be done separately from the call to the simulator, so it was not possible to determine the split in execution time between (re)compilation and simulation of the circuit. 

To compare the compilation times of OpenQL, \openqlpc{} and Qiskit, a quantum circuit was generated as in \cref{subsec:maxcut}. To limit the influence of other factors on the measurement results, the angles for each iteration were not generated using a classical optimiser, but randomly generated outside of the timing loop. The number of steps was set at 3, as mentioned before, and the number of nodes at 15. This corresponds to a circuit with a total of 330 CNOTs and rotation gates, and 15 measurement operators. For each measurement, the language (OpenQL, \openqlpc{} or Qiskit) and the total number of iterations are randomly selected. The compile time is measured from the start of any language-specific preamble and until all iterations have completed. For the iterations a simple loop is used that compiles the circuit with different angle values for the parameters. %with different angles each time, etc. Also mention that its just a for-loop?

Timing was done for a total of 1, 2, 4, 6, 8, 10, 12, 14 and 16 iterations, with and without writing of quantum assembly language to an output file. Although the circuit that is used in these tests comes from the MAXCUT benchmark, no classical optimiser is used for this set of tests. This way, the number of iterations is not dependent on any unknown factors, and the compilation time does not include the runtime of the optimiser. The Qiskit optimisation level was set to 1, which corresponds "light optimisation" \cite{misc:qiskitdocumentation}. For OpenQL, the "RotationOptimizer" pass was added to the compiler.

To determine whether the improvements made had an impact on any real applications, a full implementation of MAXCUT was used. The circuit and the parameters are the same as in the previous set of tests, but now a classical optimiser was added to the loop to determine the parameter values for the next iteration. The circuits were also run on a simulator, and the measurement results coming from the simulator are used as input for the optimiser. This makes the total execution flow as in \cref{fig:update_parameter_loop}. The tests were performed for regular graphs with between 3 and 8 nodes, with 3 steps to the algorithm as before. Tests were performed interleaved where possible, and number of nodes was randomised for each trial. For each run of the MAXCUT benchmark, the number of function evaluations was limited to 100, and any runs that reached convergence earlier where discarded. Therefore the total number of function evaluations for each language is 100, although the number of circuit compilations and simulations can be lower. This is because the loop is aborted preemptively if the optimiser generates negative angles or angles bigger than 2$\pi$. The effect this has on the results is expected to be small, and it should be the same for each language so does not influence the comparisons made. 

\section{Results} \label{sec:results}

\begin{figure}[h]
    \centering
    % \begin{subfigure}[b]{\linewidth}
        \includegraphics[width=\linewidth]{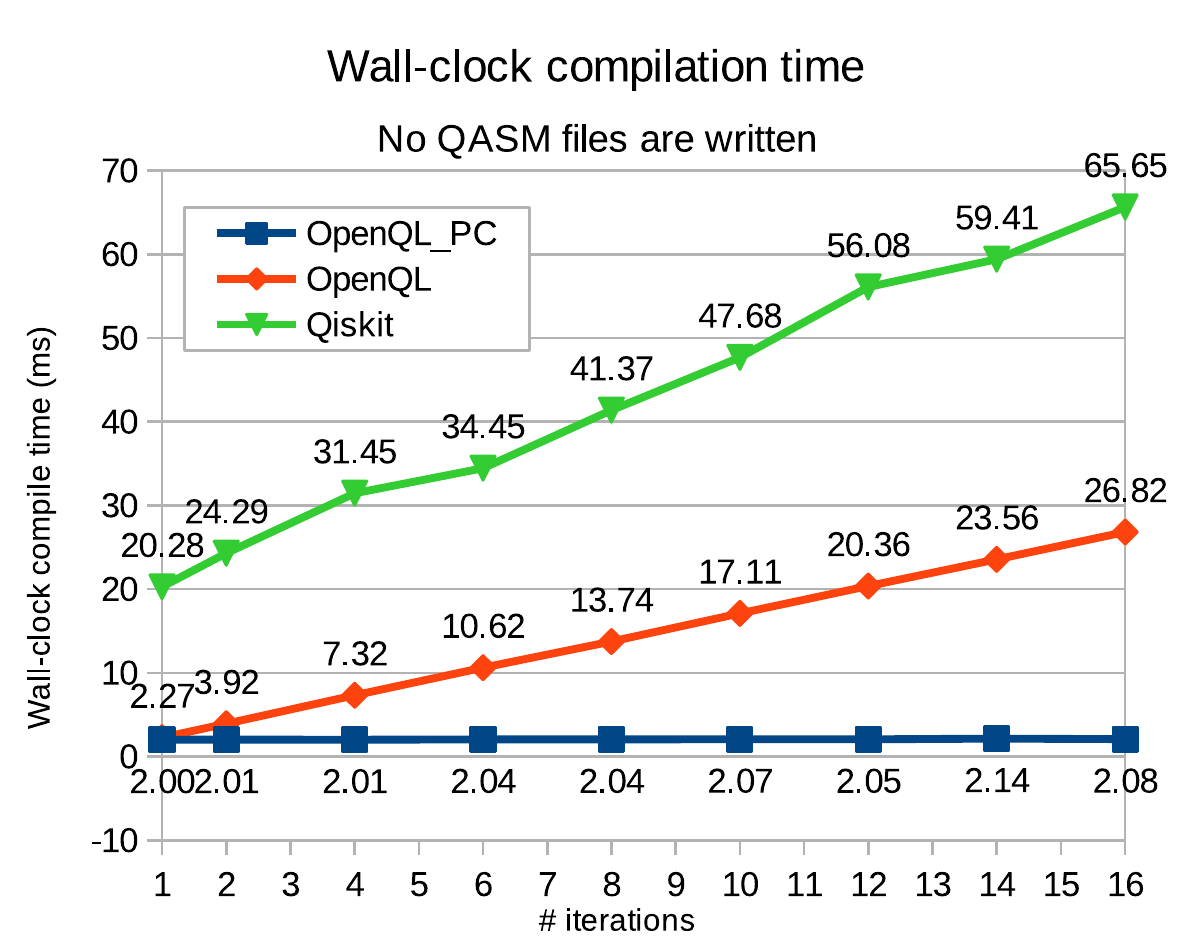}
    \caption{Wall-clock compilation time for a MAXCUT circuit with 15 qubits with varying number of iterations for Qiskit, OpenQL and \openqlpc{}, when no QASM files are written} \label{fig:new_comp_no_qasm}
\end{figure}

\begin{figure}[h]
\centering
    \begin{subfigure}[b]{\linewidth}
        \includegraphics[width=\linewidth]{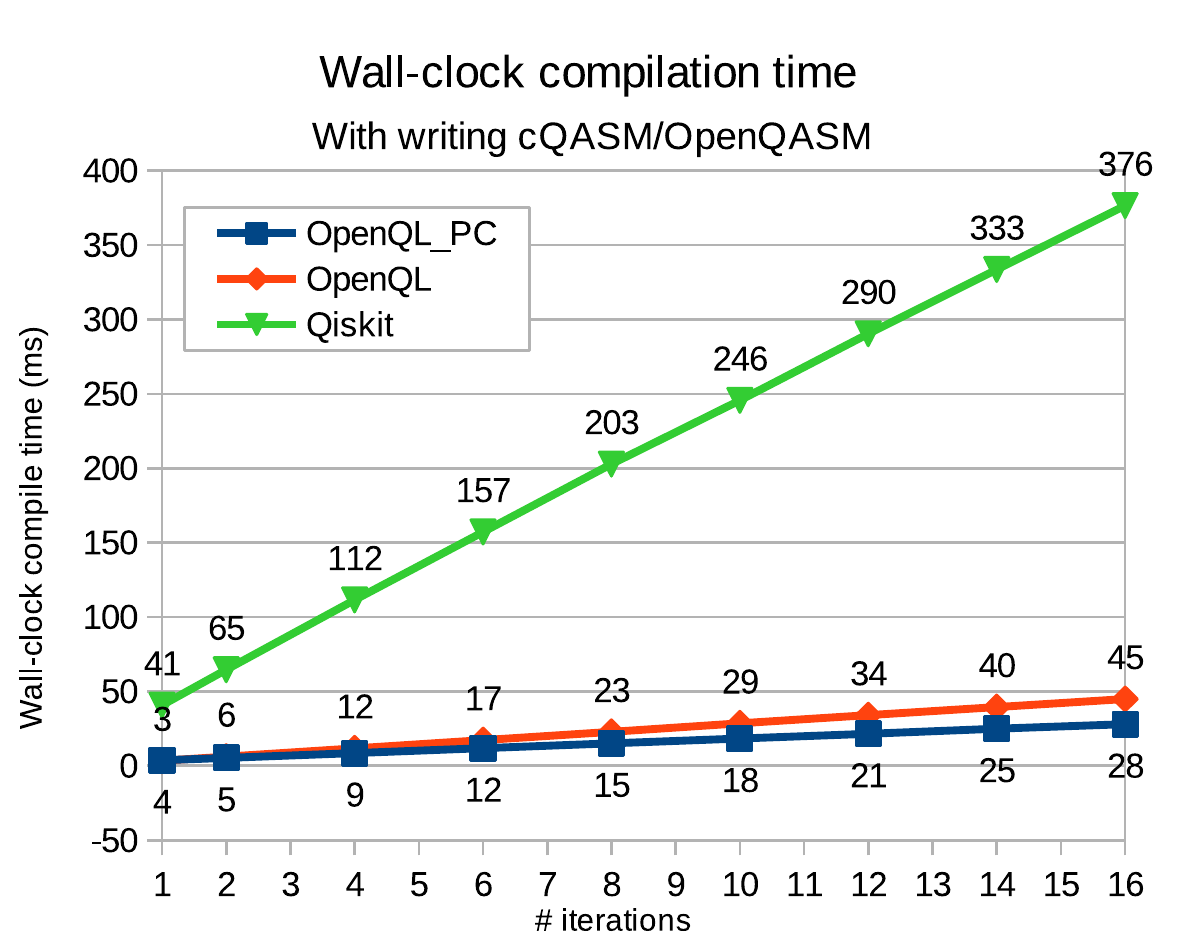}
    \caption{Wall-clock compilation time}
    \label{fig:abs_comp_yes_qasm_no_opt}
    \end{subfigure}
    \begin{subfigure}[b]{\linewidth}
        \includegraphics[width=\linewidth]{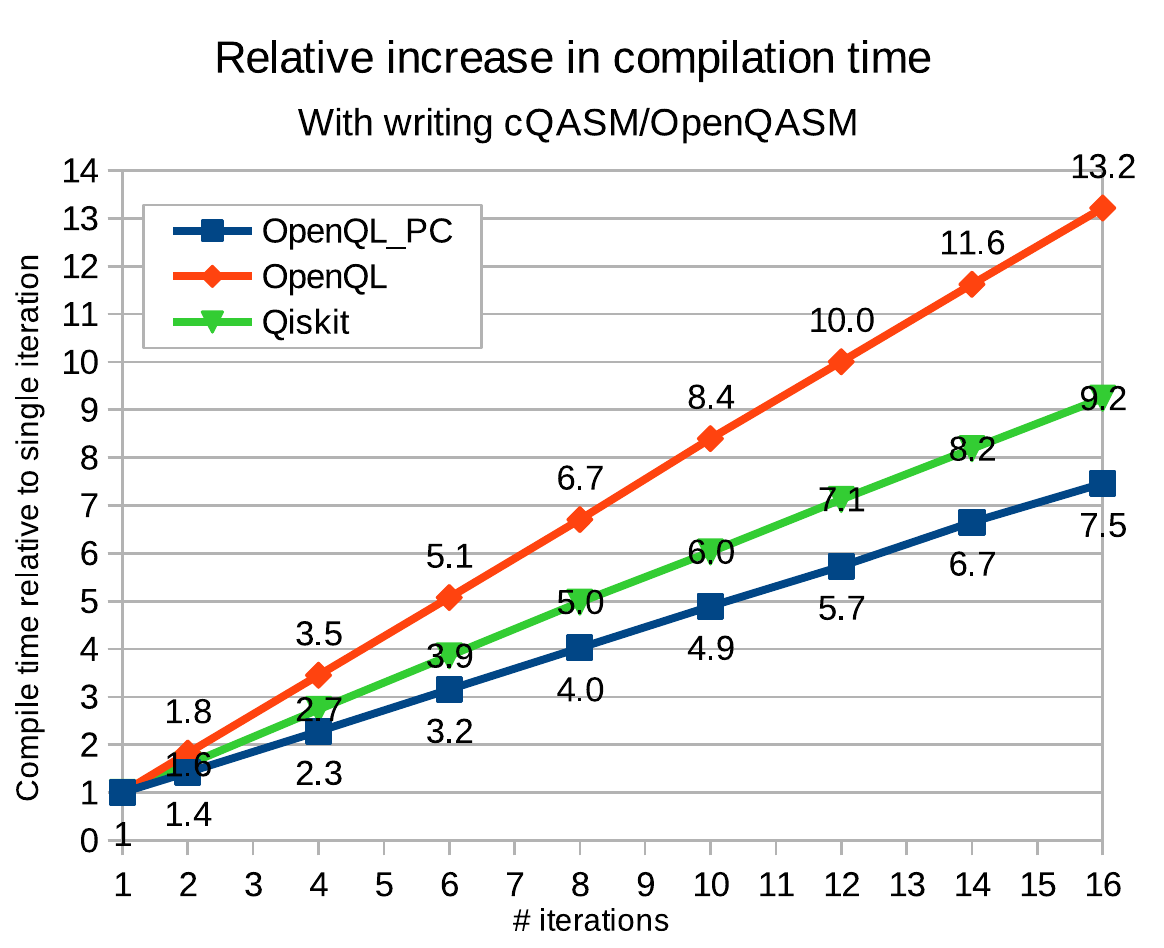}
    \caption{Compilation time relative to the compilation time of a single iteration}
    \label{fig:rel_comp_yes_qasm_no_opt}
    \end{subfigure}
    \caption{(Wall-clock) compilation time for a MAXCUT circuit with 15 nodes with varying number of iterations for Qiskit, OpenQL and \openqlpc{}, with OpenQL writing cQASM files and Qiskit writing the circuit to an OpenQASM file} \label{fig:new_comp_yes_qasm}
\end{figure}

The results for the first set of tests outlined in \cref{subsec:experimentalsetup} (without classical optimiser) can be found in \cref{fig:new_comp_no_qasm,fig:new_comp_yes_qasm}. %The raw data can be found in \cref{tab:wall-clock-no-qasm,tab:wall-clock-yes-qasm}. 
\cref{fig:new_comp_no_qasm,fig:abs_comp_yes_qasm_no_opt} show the wall-clock time in seconds that it cost to compile for the different numbers of iterations. \cref{fig:rel_comp_yes_qasm_no_opt} shows these measurements relative to the time of a single iteration (i.e., divided by the time it costs to do just one iteration). 

As can be seen in \cref{fig:new_comp_no_qasm}, \openqlpc{} does not perform any additional work beyond the first circuit compilation when there is no QASM output. Therefore, there is (almost) no increase in wall-clock compilation time when the circuit is compiled multiple times. The reason for this is that the parameters are only replaced by their numerical values when they are required, which is only when QASM is generated. 

When QASM output is generated, both \openqlpc{} and OpenQL are considerably faster than Qiskit, as can be seen in \cref{fig:abs_comp_yes_qasm_no_opt}. However, considering the relative increase in compilation times in \cref{fig:rel_comp_yes_qasm_no_opt} shows that OpenQL handles repeat compilations worse than either Qiskit or \openqlpc{}, since it is not optimised for this operation. In OpenQL, it takes 13 times as long to do 16 iterations. This is because of setting up the circuit and the compiler, which are done for each iteration instead of only once for the complete run in \openqlpc{}. The time that is saved by \openqlpc{} for subsequent iterations can be seen in both absolute and relative decrease  in compile time. OpenQL and \openqlpc{} take almost the same amount of time for a single compilation, but for each subsequent iteration, there is more and more time saved by \openqlpc{}. 

Looking at the absolute compile times shows that Qiskit is an overall much slower compiler than OpenQL or \openqlpc{}. Inspecting the relative cost of compiling for multiple iterations, Qiskit shows some level of optimisation for this type of repeated compilations. %If it did not, its plotted line would lie closer to the one of OpenQL. 
Still these optimisations are not as effective as the ones implemented %But more of the compilation is repeated for each iteration than 
in \openqlpc{}, which is faster both in absolute terms and in relative cost of compilation.

\begin{figure}[h]
\centering
        \includegraphics[width=\linewidth]{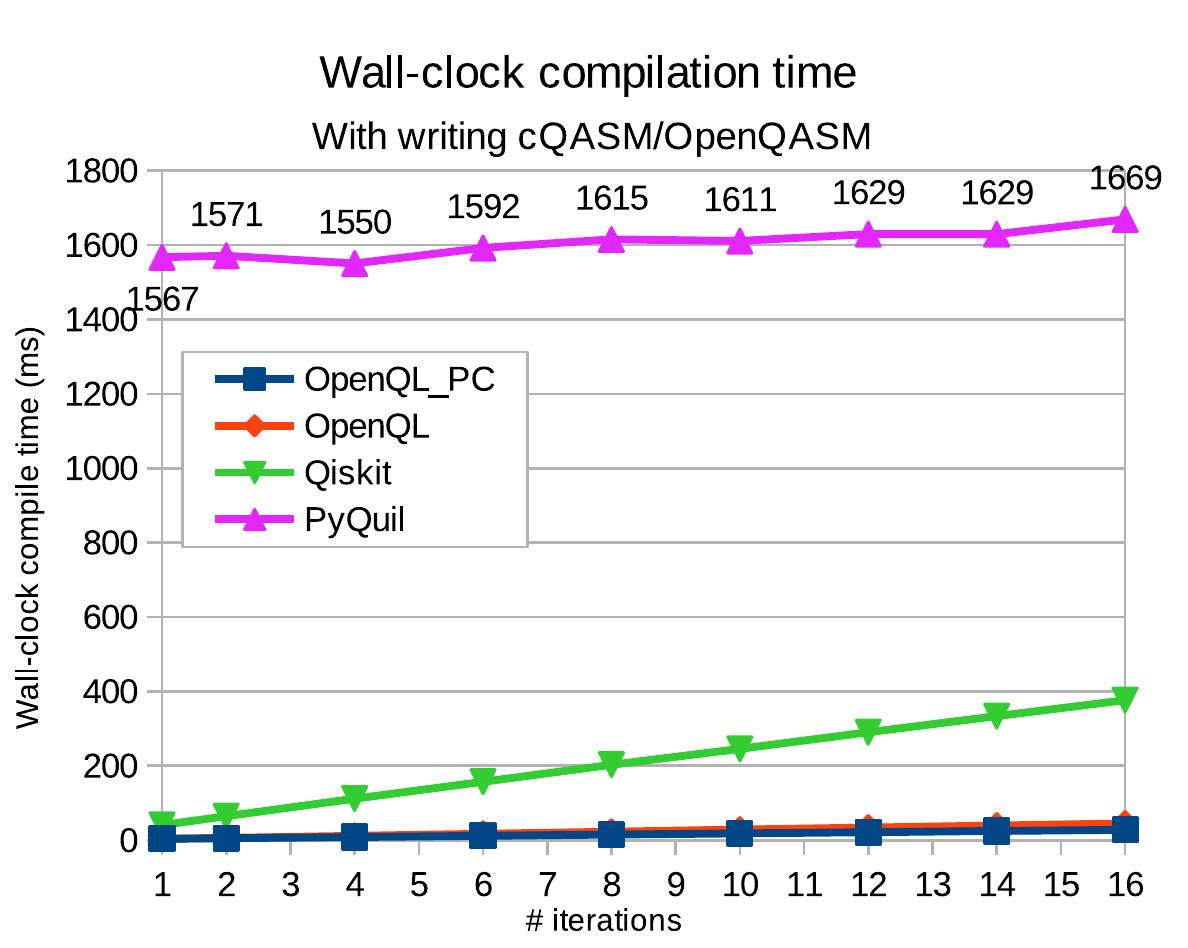}
    \caption{(Wall-clock) compilation time for a MAXCUT circuit with 15 qubits with varying number of iterations for Qiskit, OpenQL, \openqlpc{} and PyQuil, with generation of QASM files} \label{fig:abs_comp_yes_qasm_also_pyquil}
\end{figure}

Inspecting the compilation time of PyQuil, %it seems that 
\cref{fig:abs_comp_yes_qasm_also_pyquil} shows that consecutive iterations do not incur much additional time. %, as can be seen in \cref{fig:abs_comp_yes_qasm_also_pyquil}. 
However, the compile times PyQuil need are 10x to 100x higher than that of the other languages. This is due to the fact that PyQuil has to be compiled by running a separate compiler on a virtual machine in server mode, which are called from the main python program.

% The compilation time results for the MAXCUT problem in OpenQL are plotted in \cref{fig:maxcut_comp_time_openql_w_wo}. 

\begin{figure}[th]
    \centering
    \includegraphics[width=\linewidth]{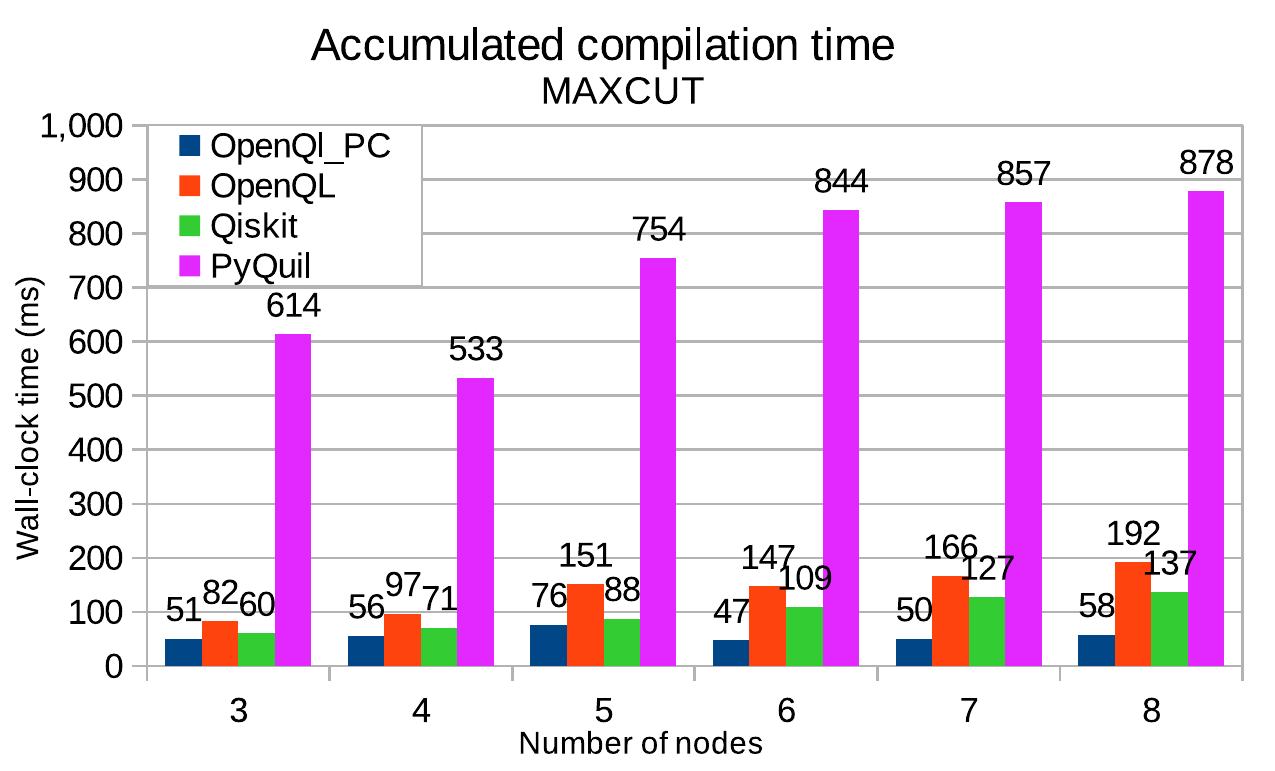}
    \caption{Accumulated compilation time for the MAXCUT benchmark with 100 iterations for graphs with 3 to 8 nodes, for \openqlpc, OpenQL, Qiskit and PyQuil}
    \label{fig:maxcut_comp_time_openql_w_wo_qiskit}
\end{figure}

In \cref{fig:maxcut_comp_time_openql_w_wo_qiskit}, we also inspect the accumulated compilation times for running MAXCUT with a classical optimiser for graphs of different numbers of nodes, for \openqlpc{}, OpenQL, Qiskit and PyQuil. \openqlpc{} has the shortest total compile time, second is Qiskit, then OpenQL, and the slowest by an order of magnitude is PyQuil. These benchmark runs were done with fewer nodes and more iterations (namely 100 iterations) than the earlier tests, and it is interesting to see that OpenQL has overtaken Qiskit in compile time. Although Qiskit takes considerably longer to compile a circuit once, as noted before, the time that is saved for each subsequent iteration results in a total shorter time spent compiling. \openqlpc{} combines the already fast compile times of OpenQL with efficiently handling iterations, and as a result is much faster than all other tested options. 

\begin{figure}[th]
    \centering
    \includegraphics[width=\linewidth]{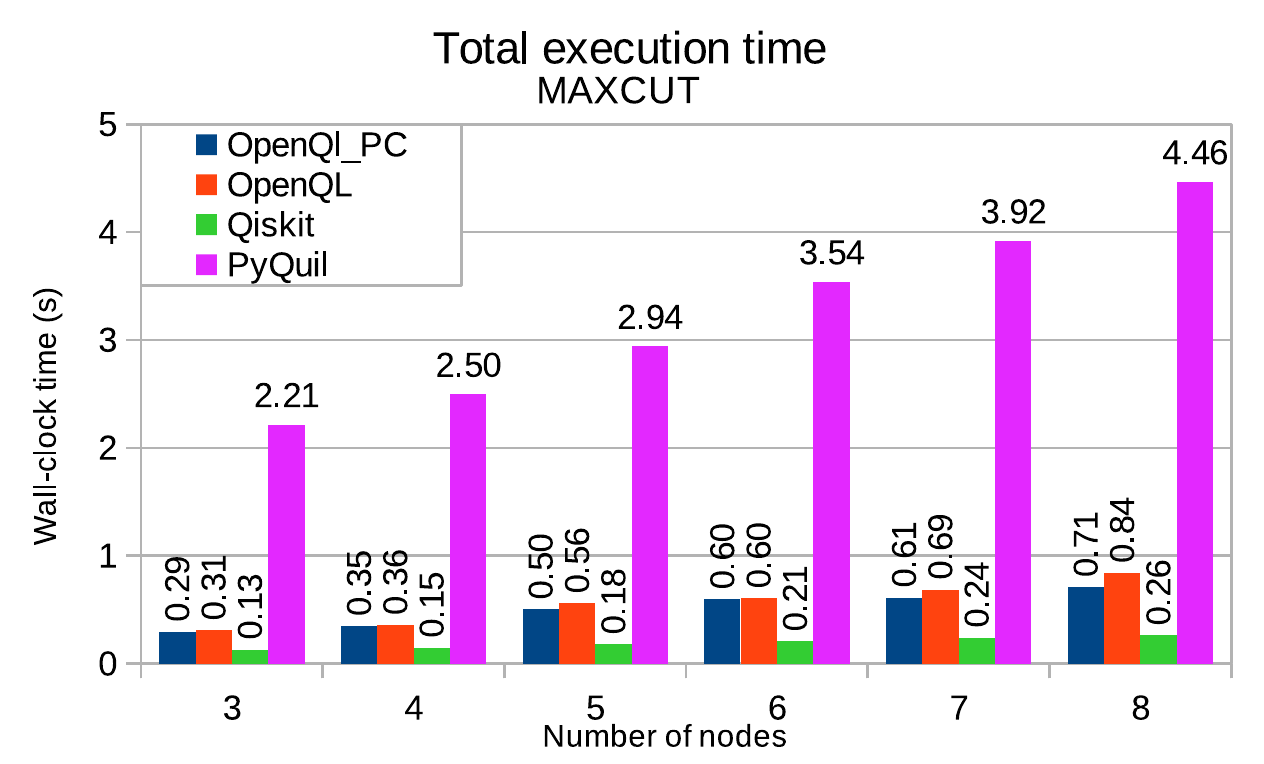}
    \caption{Total execution time for the MAXCUT benchmark with 100 iterations for graphs with 3 to 8 nodes, for \openqlpc{}, OpenQL, Qiskit and PyQuil}
    \label{fig:maxcut_total_time_openql_w_wo_qiskit}
\end{figure}

\cref{fig:maxcut_total_time_openql_w_wo_qiskit} shows the total execution time of the MAXCUT benchmark, which includes both compilation time as well as simulation time. The figure shows that total execution times are up to 50x higher than compilation times (\cref{fig:maxcut_comp_time_openql_w_wo_qiskit}).
%The total execution time of the complete benchmark, which is plotted in \cref{fig:maxcut_total_time_openql_w_wo_qiskit}, shows that the total compile time is in all cases a small part of the total execution time. 
Therefore, a big part of the difference in execution time between the programming languages is due to the difference in performance between the corresponding simulators. Still, the same trend can be seen, where PyQuil takes the longest time by far, and both OpenQL and \openqlpc{} have similar performance, since \openqlpc{} and OpenQL both use the QX simulator. %, the difference in total execution time is solely due to time saved during the compilations. 
The figure also shows that the Qiskit Aer simulator is the fastest of the tested options. 

In summary, the results show that the improvements made for \openqlpc{} result in a clear speedup compared to OpenQL. Compared to Qiskit and PyQuil, \openqlpc{} has the fastest compile times for the MAXCUT benchmark. However, when looking at the total execution time, the fast simulator time of Qiskit Aer results in the shortest total execution time of the benchmark. This can be partly explained by the large communication time between \openqlpc{}/OpenQL and the QX simulator, which requires writing and reading of a QASM file for every iteration. This is especially apparent in the MAXCUT benchmark, which has a lot of iterations for relatively short circuits, which results in a lot of read/write operations compared to circuit compilation(s). Between the Qiskit compiler and the Qiskit Aer simulator, however, the circuit can be passed directly. % And I'd love to cite some kind of future development of OpenQL/QX to say that that will improve in the future, but alas.

%% file: 05_Conclusion.tex
\section{Conclusion} \label{sec:conclusion}
In this paper, we introduced \openqlpc{}, an efficient approach to parametric compilation for hybrid quantum-classical algorithms, and implemented it in the OpenQL programming framework. \openqlpc{} is designed to be modular, scalable, usable, future-proof and fast. We compared wall-clock compilation time of \openqlpc{} with OpenQL, Qiskit and PyQuil. The total compilation time was measured using the MAXCUT benchmark from \cite{art:qpack}. 

%For the first set of tests, a quantum circuit was generated for the MAXCUT benchmark with 15 nodes and 3 steps. This circuit was compiled between 1 and 16 times, with different angle parameters for each iteration. %For one iteration, \openqlpc \ and OpenQL compile the circuit in an equal amount of time, as expected. Without QASM output, \openqlpc \ does not do any compile work for later iterations. With QASM output, the time saved by \openqlpc \ starts to show, and by 16 iterations it is almost twice as fast as OpenQL. 
Experimental results show that compared to other programming languages, %compilation in OpenQL is already much faster for these circuits, even without the improvements of \openqlpc.  Looking at Qiskit, OpenQL and \openqlpc \ showed that again,   
total compile time of OpenQL and \openqlpc{} is between 10 and 20 times faster than Qiskit, respectively. PyQuil had the slowest time at about 60x slower than \openqlpc{}. In addition, comparing compile times of multiple compile iterations relative to single iterations shows that \openqlpc{} is the fastest, followed by Qiskit which is 1.2x slower, while OpenQL took the most time being 1.8x slower than \openqlpc{}. %, where compilation of 16 iterations took 9.2 times as long as compilation of 1 iteration for the former, and 13.2 times for the latter, and \openqlpc{} had both of them beat with taking only 7.5 times as long. Compilation times for PyQuil where more than a hundred times as long as those for \openqlpc{}, regardless of the number of iterations. 

% add second set of tests also
The MAXCUT benchmark was also implemented in its entirety, including a classical optimiser and simulations. These tests, with shorter circuits but more iterations than before, show that the improvements made in \openqlpc{} result in a decrease of 40 to 70\% in (accumulated) compilation time. This makes \openqlpc{} faster than all other tested languages. For the total execution (run)time of the MAXCUT benchmark, the performance of the simulators used has more influence than the compile times of the languages. The simulator used with PyQuil is still the slowest option by far, but the Qiskit Aer simulator is 2 to 3 times as fast as the QX simulator used by OpenQL and \openqlpc{}. Even so, the faster compile time of \openqlpc{} does lead to some speed-up of the complete benchmark.

%The measurement results of the MAXCUT benchmark on OpenQL show that using the new, efficient implementation of parameters offers a clear speed-up . 
%???!add some numbers about the results???!!
It is important to note that the measurements of the complete MAXCUT benchmark show that when simulating quantum algorithms on classical computers, the simulation time can be excessively large as expected, which limits the impact of the improvement achieved by our approach. %is relatively small compared to the time required for simulation.%, the compilation time is a small improvement. 
However, the impact of our approach becomes significant when running these algorithms on real quantum devices. %In the case of execution on real quantum devices, the actual circuit execution will take much less time. For one, the reason these kinds of algorithms are considered at all is because they are very inefficient to execute on classical computers. The quantum tomography step is also completely parallelisable, both on a simulator and on real quantum devices~\cite{art:adaptivevariationalalgnature}. 
Furthermore, %simulation of larger circuits requires multiplication of exponentially larger vectors and matrices, which is obviously not the case for execution on real quantum devices. Finally, 
a single compilation of a quantum circuit is projected to become more computationally expensive as more sophisticated mapping, optimisation and error-correcting algorithms are created and implemented, which will further increase the cost of repeated compilations in hybrid algorithms and increase the impact of our approach. 

%In conclusion, efficient compilation of parameters can significantly decrease compilation time in hybrid algorithms. For longer circuits, more qubits and more sophisticated compiler-level optimisations, this benefit will only increase. % For execution on current quantum simulators, the compilation does not affect total execution time much, but quantum circuits will (hopefully) not be simulated forever and compilation costs will only increase. So therefore this improvement can contribute to the achievement of actual quantum supremacy.

%% file: main.bib
@article{art:googlequantumsupremacy2019,
title	= {Quantum Supremacy using a Programmable Superconducting Processor},
author	= {Frank Arute and Kunal Arya and Ryan Babbush and Dave Bacon and Joseph Bardin and Rami Barends and Rupak Biswas and Sergio Boixo and Fernando Brandao and David Buell and Brian Burkett and Yu Chen and Jimmy Chen and Ben Chiaro and Roberto Collins and William Courtney and Andrew Dunsworth and Edward Farhi and Brooks Foxen and Austin Fowler and Craig Michael Gidney and Marissa Giustina and Rob Graff and Keith Guerin and Steve Habegger and Matthew Harrigan and Michael Hartmann and Alan Ho and Markus Rudolf Hoffmann and Trent Huang and Travis Humble and Sergei Isakov and Evan Jeffrey and Zhang Jiang and Dvir Kafri and Kostyantyn Kechedzhi and Julian Kelly and Paul Klimov and Sergey Knysh and Alexander Korotkov and Fedor Kostritsa and Dave Landhuis and Mike Lindmark and Erik Lucero and Dmitry Lyakh and Salvatore Mandrà and Jarrod Ryan McClean and Matthew McEwen and Anthony Megrant and Xiao Mi and Kristel Michielsen and Masoud Mohseni and Josh Mutus and Ofer Naaman and Matthew Neeley and Charles Neill and Murphy Yuezhen Niu and Eric Ostby and Andre Petukhov and John Platt and Chris Quintana and Eleanor G. Rieffel and Pedram Roushan and Nicholas Rubin and Daniel Sank and Kevin J. Satzinger and Vadim Smelyanskiy and Kevin Jeffery Sung and Matt Trevithick and Amit Vainsencher and Benjamin Villalonga and Ted White and Z. Jamie Yao and Ping Yeh and Adam Zalcman and Hartmut Neven and John Martinis},
year	= {2019},
URL	= {https://www.nature.com/articles/s41586-019-1666-5},
journal	= {Nature},
pages	= {505–510},
volume	= {574}
}

@article{art:hybridquantumclassical2021,
author = {Endo ,Suguru and Cai ,Zhenyu and Benjamin ,Simon C. and Yuan ,Xiao},
title = {Hybrid Quantum-Classical Algorithms and Quantum Error Mitigation},
journal = {Journal of the Physical Society of Japan},
volume = {90},
number = {3},
pages = {032001},
year = {2021},
doi = {10.7566/JPSJ.90.032001},

URL = { 
        https://doi.org/10.7566/JPSJ.90.032001
    
},
eprint = { 
        https://doi.org/10.7566/JPSJ.90.032001
    
}

}

@article{art:theortofvariationalhybrid2016,
	doi = {10.1088/1367-2630/18/2/023023},
	url = {https://doi.org/10.1088/1367-2630/18/2/023023},
	year = 2016,
	month = {2},
	publisher = {{IOP} Publishing},
	volume = {18},
	number = {2},
	pages = {023023},
	author = {Jarrod R McClean and Jonathan Romero and Ryan Babbush and Al{\'{a}}n Aspuru-Guzik},
	title = {The theory of variational hybrid quantum-classical algorithms},
	journal = {New Journal of Physics}
}

@article{art:characerizingquantumsupremacyBoixo2018,
   title={Characterizing quantum supremacy in near-term devices},
   volume={14},
   ISSN={1745-2481},
   url={http://dx.doi.org/10.1038/s41567-018-0124-x},
   DOI={10.1038/s41567-018-0124-x},
   number={6},
   journal={Nature Physics},
   publisher={Springer Science and Business Media LLC},
   author={Boixo, Sergio and Isakov, Sergei V. and Smelyanskiy, Vadim N. and Babbush, Ryan and Ding, Nan and Jiang, Zhang and Bremner, Michael J. and Martinis, John M. and Neven, Hartmut},
   year={2018},
   month={4},
   pages={595–600}
}

@inproceedings{art:QX2017,
title = "QX: A high-performance quantum computer simulation platform ",
author = "Nader Khammassi and Imran Ashraf and Xiang Fu and {Garc{\'i}a Almudever}, Carmina and Koen Bertels",
year = "2017",
doi = "10.23919/DATE.2017.7927034",
language = "English",
isbn = "978-1-5090-5826-6",
pages = "464--469",
booktitle = "Proceedings of the 2017 Design, Automation \& Test in Europe Conference \& Exhibition (DATE)",
publisher = "IEEE ",
address = "United States",
note = "Design, Automation and Test in Europe : DATE 17 ; Conference date: 27-03-2017 Through 31-03-2017"
}

@misc{misc:qiskitdocumentation,
url={https://qiskit.org/documentation/},
note={Accessed on: 10-07-2020},
author={{Qiskit Development team}},
title={Qiskit documentation},
year={2020}}

@misc{misc:pyquilgithub,
url={https://github.com/rigetti/pyquil},
title={Pyquil: Quantum programming in Python},
note={Accessed on: 24-06-2020},
author={Smith, Robert and Zeng, Will and Curtis, Spike and Rubin, Nick and Polloreno, Anthony and Karalekas, Peter and Tezak, Nikolas and Osborn, Chris and many more},
year={2020}}

@misc{misc:pyquildoc,
url={https://pyquil-docs.rigetti.com/en/stable},
title={PyQuil documention},
author={Rigetti Computing},
year={2021}}

@misc{misc:quilspec,
    title={A Practical Quantum Instruction Set Architecture},
    author={Robert S. Smith and Michael J. Curtis and William J. Zeng},
    year={2016},
    eprint={1608.03355},
    archivePrefix={arXiv},
    primaryClass={quant-ph}
}

@article{art:qcspaperrigetti,
    title = {A quantum-classical cloud platform optimized for variational hybrid algorithms},
    author = {Peter J Karalekas and Nikolas A Tezak and Eric C Peterson
              and Colm A Ryan and Marcus P da Silva and Robert S Smith},
    year = 2020,
    month = {4},
    publisher = {{IOP} Publishing},
    journal = {Quantum Science and Technology},
    volume = {5},
    number = {2},
    pages = {024003},
    doi = {10.1088/2058-9565/ab7559},
    url ={https://doi.org/10.1088\%2F2058-9565\%2Fab7559},
}

@article{avariationaleigenvaluesolver_Peruzzo_2014,
   title={A variational eigenvalue solver on a photonic quantum processor},
   volume={5},
   ISSN={2041-1723},
   url={http://dx.doi.org/10.1038/ncomms5213},
   DOI={10.1038/ncomms5213},
   number={1},
   journal={Nature Communications},
   publisher={Springer Science and Business Media LLC},
   author={Peruzzo, Alberto and McClean, Jarrod and Shadbolt, Peter and Yung, Man-Hong and Zhou, Xiao-Qi and Love, Peter J. and Aspuru-Guzik, Alán and O’Brien, Jeremy L.},
   year={2014},
   month={7}
}

@misc{art:khammassi2020openql,
      title={OpenQL : A Portable Quantum Programming Framework for Quantum Accelerators}, 
      author={N. Khammassi and I. Ashraf and J. v. Someren and R. Nane and A. M. Krol and M. A. Rol and L. Lao and K. Bertels and C. G. Almudever},
      year={2020},
      eprint={2005.13283},
      archivePrefix={arXiv},
      primaryClass={quant-ph}
}

@article{art:proglangandcompilerdesign,
author = {Chong, Frederic and Martonosi, Margaret},
year = {2017},
month = {09},
pages = {180-187},
title = {Programming languages and compiler design for realistic quantum hardware},
volume = {549},
journal = {Nature},
doi = {10.1038/nature23459}
}

@article{art:qpack,
  author    = {Koen Mesman and
               Zaid Al{-}Ars and
               Matthias M{\"{o}}ller},
  title     = {QPack: Quantum Approximate Optimization Algorithms as universal benchmark
               for quantum computers},
  journal   = {CoRR},
  volume    = {abs/2103.17193},
  year      = {2021},
  url       = {https://arxiv.org/abs/2103.17193},
  archivePrefix = {arXiv},
  eprint    = {2103.17193},
  bibsource = {dblp computer science bibliography, https://dblp.org}
}

@article{art:acomparisonofvariousclassical,
author = {Pellow-Jarman, Aidan and Sinayskiy, Ilya and Pillay, Anban and Petruccione, Francesco},
year = {2021},
month = {06},
pages = {},
title = {A Comparison of Various Classical Optimizers for a Variational Quantum Linear Solver}
}

@article{art:adaptivevariationalalgnature,
   title={An adaptive variational algorithm for exact molecular simulations on a quantum computer},
   volume={10},
   ISSN={2041-1723},
   url={http://dx.doi.org/10.1038/s41467-019-10988-2},
   DOI={10.1038/s41467-019-10988-2},
   number={1},
   journal={Nature Communications},
   publisher={Springer Science and Business Media LLC},
   author={Grimsley, Harper R. and Economou, Sophia E. and Barnes, Edwin and Mayhall, Nicholas J.},
   year={2019},
   month={7}
}

@article{art:VQEwithfewerqubits,
   title={Variational quantum eigensolver with fewer qubits},
   volume={1},
   ISSN={2643-1564},
   url={http://dx.doi.org/10.1103/PhysRevResearch.1.023025},
   DOI={10.1103/physrevresearch.1.023025},
   number={2},
   journal={Physical Review Research},
   publisher={American Physical Society (APS)},
   author={Liu, Jin-Guo and Zhang, Yi-Hong and Wan, Yuan and Wang, Lei},
   year={2019},
   month={9}
}

@article{art:HWefficientVQE,
   title={Hardware-efficient variational quantum eigensolver for small molecules and quantum magnets},
   volume={549},
   ISSN={1476-4687},
   url={http://dx.doi.org/10.1038/nature23879},
   DOI={10.1038/nature23879},
   number={7671},
   journal={Nature},
   publisher={Springer Science and Business Media LLC},
   author={Kandala, Abhinav and Mezzacapo, Antonio and Temme, Kristan and Takita, Maika and Brink, Markus and Chow, Jerry M. and Gambetta, Jay M.},
   year={2017},
   month={9},
   pages={242–246}
}

@article{art:Barahona1988AnAO,
  title={An Application of Combinatorial Optimization to Statistical Physics and Circuit Layout Design},
  author={Francisco Barahona and Martin Gr{\"o}tschel and Michael J{\"u}nger and Gerhard Reinelt},
  journal={Oper. Res.},
  year={1988},
  volume={36},
  pages={493-513}
}

@article{art:WANG2010240,
title = {Maximum cut in fuzzy nature: Models and algorithms},
journal = {Journal of Computational and Applied Mathematics},
volume = {234},
number = {1},
pages = {240-252},
year = {2010},
issn = {0377-0427},
doi = {https://doi.org/10.1016/j.cam.2009.12.022},
url = {https://www.sciencedirect.com/science/article/pii/S0377042709008309},
author = {Rui-Sheng Wang and Li-Min Wang}
}

@misc{art:qaoafarhi,
      title={A Quantum Approximate Optimization Algorithm}, 
      author={Edward Farhi and Jeffrey Goldstone and Sam Gutmann},
      year={2014},
      eprint={1411.4028},
      archivePrefix={arXiv},
      primaryClass={quant-ph}
}

@INPROCEEDINGS{art:shors,
  author={Shor, P.W.},
  booktitle={Proceedings 35th Annual Symposium on Foundations of Computer Science}, 
  title={Algorithms for quantum computation: discrete logarithms and factoring}, 
  year={1994},
  volume={},
  number={},
  pages={124-134},
  doi={10.1109/SFCS.1994.365700}}

@misc{art:grovers,
  doi = {10.48550/ARXIV.QUANT-PH/9605043},  
  url = {https://arxiv.org/abs/quant-ph/9605043},  
  author = {Grover, Lov K.},  
  title = {A fast quantum mechanical algorithm for database search},
  publisher = {arXiv},  
  year = {1996},  
  copyright = {Assumed arXiv.org perpetual, non-exclusive license to distribute this article for submissions made before January 2004}
}

@article{art:quaser,
    doi = {10.1371/journal.pone.0249850},
    author = {Sarkar, Aritra AND Al-Ars, Zaid AND Bertels, Koen},
    journal = {PLOS ONE},
    publisher = {Public Library of Science},
    title = {QuASeR: Quantum Accelerated de novo DNA sequence reconstruction},
    year = {2021},
    month = {04},
    volume = {16},
    url = {https://doi.org/10.1371/journal.pone.0249850},
    pages = {1-23},
    number = {4},

}
